\documentclass[aps,prb,twocolumn,amsmath]{revtex4}
\usepackage[dvips]{color}
\usepackage[normalem]{ulem}

\definecolor{g-blue}{rgb}{0.83,0.95,1}
\definecolor{g-yellow}{rgb}{1,1,0.7}
\definecolor{g-green}{rgb}{0.9,1,0.9}
\definecolor{green}{rgb}{0,0.6,0}
\definecolor{cyan}{rgb}{0,0.7,0.7}
\definecolor{black}{rgb}{0,0,0}
\definecolor{grey}{rgb}{0.4 ,0.4 ,0.4 }

\usepackage{bm}
\usepackage{amssymb}
\usepackage{amsfonts}
\usepackage{amsmath}
\usepackage{graphicx}
\usepackage{amsmath,bm,epsfig}
\def \ed {\end{document}}
\def\Fbox#1{\vskip1ex\hbox to 8.5cm{\hfil\fboxsep0.3cm\fbox{%
		\parbox{8.0cm}{#1}}\hfil}\vskip1ex\noindent}  

\newcommand{\Eq}[1]{Eq.\,(\ref{#1})}
\newcommand{\Eqs}[1]{Eqs.\,(\ref{#1})}
\newcommand{\Fig}[1]{Fig.\,\ref{#1}}
\newcommand{\Figs}[1]{Figs.\,\ref{#1}}
\newcommand{\Sec}[1]{Sec.\,\ref{#1}}
\newcommand{\Ref}[1]{Ref.\,\cite{#1}}
\newcommand{\Refs}[1]{Refs.\,\cite{#1}}

\def\be{\begin{equation}}
\def\ee{\end{equation}}
\def\bea{\begin{eqnarray}}
\def\eea{\end{eqnarray}}
\def\bse{\begin{subequations}}
\def\ese{\end{subequations}}

\let \nn  \nonumber

\let \= \equiv  \let\*\cdot \let\~\widetilde \let\^\widehat \let\-\overline
\let\p\partial

 \def\1{\bm1} 

\def\<{\left\langle}    \def\>{\right\rangle}
\def\({\left(}          \def\){\right)}
\def \[ {\left [} \def \] {\right ]}

\newcommand{\D}{\Delta}\newcommand{\ve}{\varepsilon}

\newcommand{\B}[1]{{\bm{#1}}}
\newcommand{\C}[1]{{\mathcal{#1}}}    
\newcommand{\BC}[1]{\bm{\mathcal{#1}}}

\renewcommand{\sb}[1]{_{\text {#1}}}  
\renewcommand{\sp}[1]{^{\text {#1}}}  
\def\Sb#1{_{\scriptscriptstyle\rm{#1}}}
\def\He4 {$^4$He~}

\begin{document}

\title{ Strong anisotropy     of  superfluid $^4$He counterflow turbulence}
\author{L. Biferale$^1$, D. Khomenko$^2$, V. L'vov$^3$, A. Pomyalov$^3$, I. Procaccia$^3$ and G. Sahoo$^4$}
\affiliation{$^1$Dept. of Physics, University of Rome, Tor Vergata,
Roma, Italy\\
$^2$  Laboratoire de Physique , D\'{e}partement de physique de l'ENS, \'{E}cole Normale Sup\'{e}rieure, PSL Research University, Sorbonne Universit\'{e}s, CNRS, 75005 Paris, France\\
$^3$Dept. of Chemical and Biological Physics, Weizmann Institute of Science, Rehovot, Israel\\
$^4$Dept. of Mathematics and Statistics and Dept. of Physics, University of Helsinki, Finland}
\begin{abstract}
We report on a combined theoretical and  numerical study 
of counterflow turbulence in superfluid $^{4}$He  in a wide range of parameters.  The energy spectra of the velocity fluctuations of both the normal-fluid and superfluid components are strongly anisotropic. The angular dependence of the correlation between  velocity fluctuations of the two components plays the key role.   A selective energy dissipation intensifies as scales decrease, with the streamwise velocity fluctuations becoming dominant. Most of the flow energy is concentrated in a wavevector plane which is orthogonal to the direction of the counterflow. The phenomenon becomes more prominent at higher temperatures as the coupling between the components depends on the temperature and the direction with respect to the counterflow velocity.
\end{abstract}

\maketitle

\section*{\label{s:intro}Introduction}
Below a critical temperature $T_\lambda\approx 2.17\,$K liquid $^4$He behaves  as a quantum fluid\cite{Donnely,2}, consisting of an inviscid superfluid, associated to the quantum 	ground state, and a gas of thermal excitations which make up the viscous normal fluid. Quantum mechanics\,\cite{Feynman} constrains the rotational motion of the superfluid in \He4 to
discrete \r{A}ngstr\"{o}m-width quantum vortex lines of fixed circulation. The thermal excitations scatter on a dense tangle of these vortices, thus inducing a mutual friction force between the  normal fluid and the superfluid.

Turbulent superfluid Helium in a channel with a temperature gradient is a  subject of extensive research for many decades\cite{Donnely,2, Vinen,Feynman,HV, BK,Vinen3,37,Schwarz88,TenChapters}.  In such a setting, so-called  ``counterflow", the normal fluid flows from the hot end of the channel to the cold end
 while the superfluid flows in the opposite direction.
Most attention was devoted so far to the measurement and the analysis of the density of vortex lines and to the mutual friction between the components.

Recent advances in the visualization techniques offer for the first time a direct access to the statistics of the velocity fluctuations of the normal fluid \cite{WG-2015,WG-2017,WG-2018} and the superfluid \cite{7,Prague1,Prague2}. It was shown that the large scale statistics of the normal fluid  in the counterflow is  very different\cite{WG-2015,WG-2017,WG-2018} from the statistics of classical fluids. The theoretical analysis\cite{decoupling,LP-2018, AnisoLetter} highlighted the importance of correlations between the superfluid and the normal fluid components, which lead to the energy spectra of both components being steeper than their classical counterparts. Moreover, we have recently shown \cite{AnisoLetter} that the direction of the mean relative velocity
 plays an important role; The correlation between the $^4$He components decays slower for eddies stretched along
the counterflow velocity. In contrast, the correlation of 
eddies, which are elongated in the orthogonal direction, decay faster, leading to their  enhanced energy loss. As a result of this directionally preferred energy dissipation, the velocity fluctuation consist mostly of the streamwise components, while most of the flow energy is concentrated in the wavevector plane orthogonal to the counterflow.

Here we consider this phenomenon further and study its consequences in further detail.
The paper is organized as follows: in the  \Sec{s:theory} we provide a sketch of a theory of counterflow turbulence with a stress on its anisotropy. In \Sec{ss:HVBK} we introduce the basic set of coarse-grained  equations for the counterflow. These are used for the theoretical analysis and for the numerical simulations.
In \Sec{ss:stat} we clarify how various approaches to the statistical description of the anisotropy energy surface are related. Next,  in  \Sec{ss:theory}, we  discuss  the physical origin of the strong spectral anisotropy in  counterflow turbulence. In  \Sec{s:DNS} we present the results of the direct numerical simulations (DNS) of the  two-fluid coarse-grained \Eqs{NSE}. The main conclusion is that the analytical predictions are confirmed. In the first subsection \Sec{ss:procedure} we discuss the simulation parameters and the numerical procedure.
Next, in \Sec{ss:1D-energy}, we use standard statistical characteristics:  one-dimensional (1D) energy spectra   and cross-correlation functions, averaged over a spherical surface of radius $k$ (i.e. over all directions of vector $\B k$) to provide an overview of spectral properties of $^4$He counterflow.
We find that at the small-$k$ regime,  the normal-fluid and superfluid velocity components are indeed well correlated. As expected,  mutual friction plays secondary role and  the spherically-averaged spectra are similar to the spectra in the $^4$He-coflow turbulence\,\cite{DNS-He4}, being only slightly steeper than the  Kolmogorov-1941 (K41) spectra of classical hydrodynamic turbulence. On the other hand, at relatively large $k$ the  fluid components are practically uncorrelated;  mutual friction provides  a leading contribution to the energy dissipation and the counterflow spectra  are  similar to those in $^3$He superfluid turbulence with the normal-fluid component at rest\cite{He3,DNS-He3}.  The spectra become strongly suppressed in comparison to K41  energy spectra.

The similarities between the inherently anisotropic counterflow energy spectra  and the isotropic spectra in the turbulent $^4$He-coflow and $^3$He are, however, superficial. To expose the differences, we discuss  in \Sec{ss:2D} the two-dimensional (2D) energy spectra  which depend, besides the wavenumber $k$, upon the angle $\theta$ between the wave-vector $\B k$ and the counterflow velocity $\B U\sb{ns}$. Here we find that  the spectra become more and more anisotropic with increasing $k$, being confined in $\B k$-space to a small range $\cos\theta < 0.1$, i.e. near the wavevectors plane which is orthogonal to $\B U\sb{ns}$.  This effect becomes stronger with increasing temperature. The tensor structure of the energy spectra,  considered in Section\,\ref{ss:tens}, is also temperature dependent: the small-scale turbulent velocity fluctuations  are dominated by only one vector component, parallel to $\B U\sb{ns}$, becoming more so at higher temperature.
Further, we  compare  several variants of differently averaged 1D spectra ( \Sec{ss:comp}) and  structure functions (\Sec{ss:SF}) to  expose other aspects of the spectral anisotropy in connection with possible experimental observations.

In \Sec{s:con} we summarize our findings: Counterflows exhibit strongly anisotropic energy distributions. The energy spectra are localized near a direction that is orthogonal to the counterflow. The phenomenon is similar to atmospheric turbulence with a strong stable  stratification or to rotational turbulence\cite{atmosTurb-Kumar,2018-AB,A5,rot1,rot2,rot3}.   On the other hand, the  tensor structures of these  two types of  quasi-2D turbulence are quite the opposite: in atmospheric turbulence the vertical component of the turbulent velocity fluctuations is suppressed by the stratification  and only the two horizontal components are dominant\cite{atmosTurb-Kumar}.  In the  counterflow turbulence  the main contribution to the turbulent energy comes from one streamwise  velocity projection, while the two  cross-steam velocity projections are strongly suppressed.  The observed phenomenon is  mild at low temperatures and becomes more prominent as the  temperature increases.  We confirm, in agreement with \Ref{WG-2018}, that the structure functions of the turbulent velocities in the counterflow  do not reflect in a quantitative manner the underlying energy spectra. However, the relative magnitude of the structure functions, measured in different directions, may qualitatively reflect the presence of the spectral anisotropy.

\section{\label{s:theory} Qualitative analysis  of anisotropic counterflow turbulence }

As we mentioned in the introduction, one of important properties of superfluid \He4 is the quantization of vorticity, which concentrates on vortex-lines of core radius $a_0\approx  10^{-8}\,$cm with fixed circulation $\kappa= h/M$$\approx 10^{-3}\,$cm$^2$/s. Here $h$ is Planck's constant  and $M$ is the mass of the \He4 atom\cite{Feynman}.  A complex tangle of these vortex lines with a typical inter-vortex distance\cite{Vinen}   $\ell\sim 10^{-4}- 10^{-2}\,$cm  is a manifestation of superfluid turbulence\cite{Feynman}.

On the large scales this type of turbulence is commonly described by the two-fluid model. The density of $^4$He $\rho$ is modelled as a mixture of two  fluid components: an inviscid  superfluid and a viscous normal fluid, with respective densities  $\rho\sb s$ and $\rho\sb n$ such that $\rho=\rho\sb s+\rho\sb n$. The fluid components are coupled by a mutual friction force \cite{Donnely,Vinen,HV,Vinen3,37,Schwarz88}.

Large-scale  turbulence in \He4 can be generated by various ways.   In mechanically driven \He4  (so-called "co-flow"), the turbulent statistics is similar\cite{SS-2012,TenChapters,BLR, Roche-new,DNS-He4} to that of classical turbulence. In this case both components move in the same direction and the mutual friction force couples them almost at all scales. On the other hand,
when a temperature gradient $\nabla T$ is imposed in a channel closed at one end, the heat flux is carried  away by the normal fluid with a mean velocity $\B U\sb n\propto \nabla T$, while
the superfluid component flows in the opposite direction with the mean  velocity $\B U\sb s$. There is no net mass flow: $\rho \sb n \B U\sb n+ \rho\sb s \B U\sb s=0$.  The counterflow velocity $\B U\sb {ns}=\B U \sb n -\B U\sb s$ creates a random vortex tangle with an energy spectrum, peaking at the intervortex scale $\ell$ and with
a close to Gaussian statistics, as  demonstrated experimentally in \Refs{7,Prague1,Prague2} and  rationalized theoretically in \Refs{BLPSV-2016,LP-2018}.
 At large enough $\B U\sb n$, the laminar flow of the normal  component  become unstable, creating large scale turbulence with  the energy spectrum dominated by ``the outer scale of turbulence" $\Delta\gg \ell$  (e.g. about half-width of the channel).
Although the particular  mechanisms of the large scale superfluid motion generation are not known in details, recent indirect experimental evidence indicate\cite{WG-2018,BLPSV-2016} that the large-scale normal fluid  motion gives rise to the superfluid turbulent motion due to the components' coupling by the mutual friction force. 

\subsection{\label{ss:HVBK}
	Coarse-grained equations for   counterflow
	$^4$He turbulence}

Our approach\cite{He4,decoupling,LP-2018} to large-scale counterflow turbulence    is based on two Navier-Stokes equations (NSE) for the velocity fluctuations of the normal fluid and superfluid components $\BC U\sb n(\B r,t)$ and $\BC U\sb s(\B r,t)$. A complication arises from the fact
that the counterflow is created in a channel and  therefore is,  in general,  inhomogeneous. However, at large enough Reynolds numbers, the flow in the center of the channel can be approximated as almost space-homogeneous\cite{Pope}. We therefore adopt a simplifying description with space homogeneity and stationarity. Further, we perform the standard Reynolds decomposition of the velocities into their mean and turbulent velocity fluctuations with zero mean:
\begin{subequations}
	\begin{eqnarray}\label{NSE}
\BC U\sb n(\B r,t)&=&\B U\sb n  +  \B u\sb n(\B r,t)\,,\  \B U\sb n =\<\BC U\sb n(\B r,t) \>\,,  \\ \nn
 \BC U\sb s(\B r,t)&=&\B U\sb s  +  \B u\sb s(\B r,t)\,,  \  \B U\sb s =\<\BC U\sb s(\B r,t) \>\ .
\end{eqnarray}
The mean velocities are taken below as externally prescribed parameters of the problem. Note that in the classical hydrodynamics, the Navier-Stokes equations are Galilean invariant and one can choose a reference system in which the constant mean velocity vanishes. In the two-fluid counterflow, there is no  such reference system and the mean velocities are necessarily present in the equations of motion for the turbulent velocity fluctuations:
\begin{eqnarray}   \label{NSEs} 
&& \hskip -0.5cm   [\frac{\p }{\p t} + (\B u\sb s+\B U\sb s)\*
\B\nabla] \B u\sb s
- \frac {\B \nabla p\sb s}{\rho\sb s }\  =\nu\sb s\,  \Delta \B u\sb s   + \B f
\sb {ns}+\B \varphi\sb s \,, 
\\  \nn
&& \hskip -0.5cm[\frac{\p }{\p t} +(\B u\sb n+\B U\sb n) \* \B
\nabla]\B u\sb n
- \frac {\B \nabla p\sb n}{\rho\sb n } = \nu\sb n\,  \Delta \B u\sb n
-\frac{\rho\sb s}{\rho\sb n}\B f \sb {ns} +\B \varphi\sb n\, ,\\ \nn
\end{eqnarray}
 These equations are coupled by the mutual friction force\, \cite{2,BK,LNV}  $\B f\sb{ns}$.  Here $\B f\sb{ns}(\B r,t)$ is a fluctuating (with zero mean) part of the total mutual friction force   $\BC  F\sb {ns}(\B r,t)$:
\begin{equation}
\BC F\sb {ns}(\B r,t)=\B F\sb {ns}   +  \B f\sb {ns}(\B r,t) \ .
\end{equation}
The pressures $p\sb n$,  $p\sb s$ in  \Eqs{NSEs} are given by
\begin{equation}
  p\sb n=\frac{\rho\sb n}{\rho }\big[p+\frac{\rho\sb s u\sb{ns}
  	^2}2\big]\,, \
p\sb s =  \frac{\rho\sb s}{\rho }\big [p-\frac{\rho\sb nu\sb{ns}
	^2 }2\big]\, ,\ \B u\sb{ns}=  \B u\sb{n}-  \B u\sb{s}\ .
\end{equation}
\end{subequations} 
The  kinematic
viscosity  of the normal fluid component is
$\nu\sb n=\eta / \rho \sb n$, where $\eta$
is the dynamical viscosity\cite{DB98}  of the normal \He4 component.  The energy sink \cite{He4} in the equation for the superfluid component,
with an  effective superfluid viscosity\,\cite{Vinen}  $\nu\sb s$,  accounts for the energy dissipation at the intervortex
scale $\ell$ due to vortex reconnections and energy transfer to  Kelvin waves.  The  random forces $\B \varphi\sb s$ and  $\B \varphi\sb n$  represent the forcing of the turbulent flow
at large scales.

The physical origin of the mutual friction is  the scattering of excitations  that
	constitute the normal fluid on the vortex lines.  Any motion of a vortex line relative to the normal fluid  results\cite{Vinen} in a
	force per unit length of the line, which can be written as
\begin{eqnarray}\label{mf}
  \B f= -\gamma_0 \B s'\times \B s' \times \big(\B U\sb {ns}  +\B u\sb {ns}\big )  +\,\gamma_0' \B s'\times \big(\B U\sb {ns}
   +\B u\sb {ns}\big ) \, .
\end{eqnarray}
 Here $\B s'$ is a unit vector along the length of the vortex, $\gamma_0$ and $\gamma'_0$ are some phenomenological  parameters.

In order to estimate the coarse-grained mutual  friction force $\B f\sb{ns}$ in \Eqs{NSEs} one needs to properly average the  microscopic \Eq{mf} for $\B f$. The result depends on the statistics of the quantum vortices  that in turn depend on the particular turbulent flow properties.   In particular,  this  procedure includes  averaging of the force $\B f$ over directions of the  orientations $\B s'$ in the vortex lines.
In the relatively simple case of rotating turbulence, the vortices are oriented mostly along the axis of rotation. In this case $\langle \B s' \rangle$ may be directly related to the direction of the superfluid vorticity $\B \omega\sb s$:	 $\langle \B s' \rangle=\B \omega\sb s/|\omega\sb s|$ . The resulting equations were named the ``HVBK equations" \cite{HV,BK}after Hall, Vinen, Bekarevich and Khalatnikov.

Clearly, the original HVBK equations are not applicable for  the superfluid turbulence without global rotation, for which $\< \B s' \>=0$. In this case, to obtain a  coarse-grained representation  one should average \Eq{mf} over ``a physically small volume" of scale $\delta$. This scale  should be chosen to be much larger than $\ell$, but still much smaller than the scale $r$ of turbulent fluctuations under consideration, $\ell \ll \delta \ll r$.  For such $\delta$, the local line orientations   $\B s'$ and $\B s' \otimes \B s'$  in \Eq{mf}  can be considered as self-averaging in space as they are almost uncorrelated with the $r$-scale fluctuations $\B u\sb{ns}(r)$, which are treated as dynamical variables.

 For co-flows a number of model expressions were suggested for  the fluctuating part of the friction force  in the form  $\B f\sb{ns}\propto \Omega\sb s \B u \sb {ns}$. Here the mutual friction frequency  $\Omega\sb s$ was modeled as a dimensional estimate, assuming underlying Kolmogorov energy spectrum for the superfluid component. Examples of such models include $\ \Omega\sb s =\alpha |\omega\sb s|$ (e.g. in \Refs{Roche-new,SCLR,SRL}), $ \Omega\sb s \alpha \sqrt {\< \omega\sb s^2\>}$(e.g. in \Refs{LNS,decoupling,DNS-He4}) and   $ \Omega\sb s= \alpha \kappa \C L$ (e.g. in   \Ref{2,WG-2018,DNS-He4,LP-2018}).
	Here $\alpha$ is the dimensionless mutual friction parameter related\cite{Donnely} to $\gamma_0$ as $\alpha \rho\sb s\kappa= (1+\alpha^2)\gamma_0$, $\C L$ is the vortex line density.

In counterflows the dynamics of the vortex tangle is dominated  by the stretching of the vortex lines by the counterflow velocity and by their  reconnections.
Based on experiments in narrow slits, Gorter and Mellink \cite{GM49} proposed to couple the equations of motion for the components' velocities by the mutual friction force of a phenomenological form  $\C F\sb {ns}= A\, \rho\sb n \rho\sb s  \,\C U\sb{ns}^3$, where $A$  is a temperature dependent constant. This form was later refined by Vinen\cite{Vinen3} for homogeneous turbulence and an isotropic vortex tangle as $\BC F\sb{ns}= G\,\BC U\sb{ns}$, $G= A\, \rho\sb n \rho\sb s \, \C U\sb{ns}^2$. Taking into account the relation between the vortex line density and the counterflow velocity in the steady-state isotropic tangle, it can be further rewritten\cite{2} as $\BC F\sb{ns}= \frac23\, \alpha \kappa \C L \, \BC U\sb{ns}$.
	
	The tangle anisotropy with respect to the direction of $\B U\sb {ns}$  can be described by the  Schwarz's  indices\cite{Schwarz88}:
\begin{subequations}\label{MF1}
	\begin{eqnarray}\label{MF1a}
I_\|=\frac 1 {L\sb{tot}} \int_{\C C} \big[1- (\B s'\cdot \B {\hat r_\|} )^2  \big ]d \xi= \<[ 1- (\B s'\cdot \B {\hat r_\|} )^2 \>_{_{\C C}} , ~~~   \\ \label{MF1b}
I_\perp =\frac 1 {L\sb{tot}} \int_{\C C} \big[1- (\B s'\cdot \B {\hat r_\perp} )^2  \big ]d \xi= \<[ 1- (\B s'\cdot \B {\hat r_\perp} )^2 \>_{_{\C C}}  . ~~
\end{eqnarray}
Here $L\sb{tot}$ is the total vortex length in the whole vortex configuration $\C C$ over which integrals are taken and  $\B {\hat r}_\| $ and $\B {\hat r}_\perp$ are unit vectors in the directions parallel and perpendicular to $\B U\sb{ns}$ respectively. Using \eqref{MF1a} and \eqref{MF1b}, the mutual friction force may be written as
\begin{eqnarray} \label{MF1c}
\B F\sb{ns}&=&  \alpha \kappa \C L   I_\|   \B U\sb {ns}\,, \\ \label{MF1d}
   \B f\sb{ns}(\B r,t)&=&  \alpha \kappa \C L \big [  I_\|   \B u^\|\sb {ns}(\B r,t)  + I_\perp\B u^\perp \sb {ns}(\B r,t)\big ] \ .
\end{eqnarray}
\end{subequations}
Notably, the second non-dissipative term in \Eq{mf}  $\propto \gamma_0'$, vanishes by symmetry  and  does not contribute \cite{Schwarz88} to the averaged quantity $\BC F\sb{ns}$.  The  mean mutual friction force  $\B F\sb{ns}$, \eqref{MF1c}, found earlier by Schwarz \cite{Schwarz88}, enters into the equations for the mean velocities $\B U\sb n$ and $\B U\sb s$, which we do not discuss. The fluctuating part \eqref{MF1d} of the mutual friction  force  $\B f\sb{ns}(\B r,t)$  enters \Eqs{NSEs}.  The vector components of turbulent counterflow velocity fluctuations
 $\B u^\|\sb {ns}$ and $\B u^\perp \sb {ns}$ are oriented   in the directions of  $\B {\hat r}_\| $ and $\B {\hat r}_\perp$ respectively.
	
The definitions \Eq{MF1} do not take into account that the turbulent intensity of the normal fluid velocity $\sqrt{\< |\B u\sb n|^2 \>} /U\sb {ns}$ in the counterflow turbulence  is not very small  and can  reach\cite{WG-2017} values of about  $0.25$. So, strictly speaking,  in \Eq{MF1a} and \eqref{MF1b} the directions  $\B {\hat r}_\| $ and $\B {\hat r}_\perp$,  should be taken  along and orthogonal to the total counterflow velocity $\BC U\sb{ns}$. In addition, one should take into  account  the space-time dependence of $\C L$ in \Eq{MF1d}.  However,  since the leading contribution   to  $\sqrt{\< |\B u\sb {ns}|^2 \> }$ originates from the outer scale of turbulence $\Delta\gg \ell$, for the motions of the scale $\ell$ this correction  effectively  adds to $U\sb{ns}$ and we can neglect the influence of the velocity fluctuations of the scale $\B r$ on the vortex line density, replace an  average of the product by the product of averages and consider $\C L$ as constant.	
The equation \eqref{MF1d} may be identically rewritten as
	\begin{eqnarray} \label{MF2c}
\B f\sb{ns}(\B r,t)&=& \Omega\sb s  \Big[ \B u \sb {ns}  + \frac{ I_\| -I_\perp}2 (   2 \B u^\|\sb{ns}-\B u^\perp\sb{ns}) \Big]  \,,\\ \nonumber
\Omega\sb s&=&\frac 23 \alpha \kappa \C L \,, \quad \B u\sb 	{ns}=\B u ^{||}\sb {ns} + \B u^\perp\sb{ns}\, .
\end{eqnarray}
Taking the numerical values of $ I_\|$ and $I_\perp$  (see e.g. Tab. IV for $T=1.6\,$K in \Ref{Kond}) we see  that
$   ( I_\perp-I_\|)/2 \approx 0.05$. This means that with a reasonable accuracy of about $(5-10)$\% we can neglect the anisotropy term in \Eq{MF2c} and use the simple form $\B f\sb{ns}(\B r,t)=  \Omega\sb s \B u\sb {ns}(\B r,t) $  as a good  approximation for $\B f\sb{ns}$ even for counterflow turbulence.  Furthermore, in this paper we do not consider the dependence of $\C L$ on the flow parameters  and use the mutual friction frequency $\Omega\sb s$ as a prescribed external control parameter. It can be estimated or measured for each particular flow conditions.

\subsection{\label{ss:stat}Statistical characteristics of anisotropic turbulence }

 \subsubsection{\label{ss:SF}Velocity correlation function}
A useful characterization of homogeneous superfluid \He4 turbulence is furnished by the three-dimensional (3D) correlation functions of the normal- and superfluid turbulent velocity fluctuations in $\B k$-representation:
 \begin{subequations}\label{def-E}
 	\begin{eqnarray}\label{def-Ea}
(2\pi)^3\delta(\B k -\B k') \C E^{\alpha\beta}  _{ij} (\bm k)&=&\<  v _i^\alpha(\bm k)   v_j^{* \beta} (\bm k')\>\,,~~~~\\ \label{def-Eb}
 \C E_{j}^{\alpha\beta}(\bm k)=  \C E_{jj}^{\alpha\beta}(\bm k)\,, \quad  \C E_{ij}(\bm k)&\=& \sum_{\alpha=x,y,z}  \C E^{\alpha\alpha}_{ij}(\bm k)   \ . ~~~
\end{eqnarray}
Here  $\delta(\B k -\B k')$ is a 3D Dirac's  delta function,  $\B v_j(\B k)$ is the Fourier transform  of  $\B u_j(\B r)$

	\begin{eqnarray}\label{def-Fc}
	\B v_j(\B k)&=&\int \B u _j(\B r)\, \exp(-i\B k\cdot \B r)\, d\B r\,,\\ \label{def-Fy}
	\B u_j(\B r)&=&\int \B v _j(\B k)\, \exp(-i\B k\cdot \B r)\, \frac{d\B k}{(2\pi)^3}\,,,
\end{eqnarray}
the indices $\alpha, \beta= \{x,y,z\}$ denote  Cartesian coordinates,
the subscripts ``$_{i,j}$"  denote  the normal
(${i,j}=$n)  or  the superfuid (${i,j}=$s) fluid components and  $^*$ stands for complex conjugation.  In the rest of the paper,  we denote the trace of any tensor
according to $\C E_{ij}(\bm k)= \sum_{\alpha}  \C E^{\alpha\alpha}_{ij}(\bm k) $ .  The correlation function $\C E^{\alpha\beta}  _{ij} (\bm k)$ and the Fourier transform\,\eqref{def-Fc} are  defined such that the kinetic energy density per unite mass $ E_j$ reads
\begin{equation}\label{def-Fd}
    E_{j }= \frac12   \< |\B u _j(\B r)|^2 \>=   \frac 12  \int  \C E _{jj} (\bm k) \frac{d^3 k}{(2\pi)^3}\ .
	\end{equation}
	 \end{subequations}
  The  dimensionality of the   energy density is   $[  E_{jj}]=$cm$^2$/s$^2$,  while the  dimensionality of  the 3D energy spectra is $[\C E_{jj}]=$cm$^5$/s$^2$.

 Due to the presence of a preferred direction (the counterflow velocity), the resulting turbulence has an axial symmetry around that  direction. Accordingly,   $\C E_{ij}^{\alpha\beta}(\B k)$ \  depends only on two     projections $k_\|$ and $k_\perp$  of the wave-vector $\B k$:    $\B k_\|\= \B U\sb{ns} (\B k\cdot \B U\sb{ns})/ U\sb{ns}^2$ and $\B k_\perp \perp \B U\sb{ns}$, being independent of the angle $\varphi$ in the $\perp$-plane, orthogonal to $\B U\sb{ns}$: $\C E_{ij}^{\alpha\beta}(\B k)\Rightarrow \C E_{ij}^{\alpha\beta} (k_\|,k_\perp)$.
This allows us to define a two-dimensional (2D) object $ E^{\alpha\beta}  _{ij} (k_\|,k_\perp)$  that still contains all the information about 2$\sp{nd}$-order statistics of the counterflow turbulence:
 \begin{subequations}\label{def-F} 
 	\begin{eqnarray}\label{def-Fa}
  && \hskip - .7cm ~ E^{\alpha\beta}  _{ij} (k_\|,k_\perp) \=    \frac{k_\perp}{4\pi^2} \C E^{\alpha\beta}  _{ij} (k_\|,k_\perp)\ .
	\end{eqnarray}
 Another way to represent the same information is to introduce  a polar angle $\theta=\angle ( \B k, \B U\sb{ns}) $, to represent the wavevector length as $k=\sqrt{k_\|^2+k_\perp^2}$  and to define a  2D object $\tilde  E^{\alpha\beta}  _{ij} (k,\theta)$ in spherical coordinates:
\begin{eqnarray}
&&\hskip - 1cm   \tilde E^{\alpha\beta}  _{ij} (k ,\theta) \=  \frac{k   \C E^{\alpha\beta}  _{ij} (\B k)}{4\pi^2}= \frac{k }{4\pi^2} \C E^{\alpha\beta}  _{ij} (k\cos\theta,k\sin\theta)  .  \label{def-Faa}
 \end{eqnarray}
 \end{subequations}
The dimensionality of 2D energy spectra and correlation functions is $[~ E^{\alpha\beta}  _{ij}  ]= [~\tilde E^{\alpha\beta}  _{ij}  ] $=cm$^4$/s$^2$.

A more compact but less detailed information on the statistics of turbulence is provided by a set of  one-dimensional (1D) energy spectra. The most traditional are the 1D ``\textit{spherical}" energy spectra
  and the cross-correlation function $^{\bullet\!}E_{ij}^{\alpha\beta}(k)$, averaged over a spherical surface of radius $k$:

\begin{subequations}
	\begin{eqnarray}   \nn
&&\hskip -.8 cm ~^{\bullet\!} E_{ ij}^{\alpha\beta} (k)=  \int \C E_{ij}^{\alpha\beta}(k_\|,k_\perp)\delta \big (k-\sqrt{k_\|^2+k_\perp^2 }\, \big)  \frac{k_\perp d k_\perp dk_\|}{4\pi^2} \\ \label{Esp}
 &=&\int  E_{ij}^{\alpha\beta}(k_\|,k_\perp)\delta (k-\sqrt{k_\|^2 +k_\perp^2 })    d k_\perp dk_\| \\ \nn
&=&  k \int_{-1}^1 ~ \tilde  E_{ij}^{\alpha\beta} (k, \theta) d \cos \theta \ .
\end{eqnarray}
In the isotropic case, when  $\C   E_{ij}^{\alpha\beta}$ depends only on $k=\sqrt{k_\|^2 +k_\perp^2 } $, this representation simplifies to a well know relationship
\begin{equation}\label{EspC}
^{\bullet\!}E_{ij}^{\alpha\beta} (k)=  \frac{k^2}{2\pi^2}\C  E_{ij}^{\alpha\beta} (k)\,, \  \mbox{for spherical symmetry. }
\end{equation}

Further information about the anisotropy of the $2\sp{nd}$-order statistics is obtained by comparing the spherical 1D-spectra  $  ^{\bullet\!}E_{ij}(k)$  with a set of 1D spectra averaged differently.  A natural choice are spectra averaged
 over a cylinder of radius $k_\perp$ with the axis oriented along $\B k_{\|}$. This results in the \textit{ cylindrical} 1D spectra
\begin{eqnarray}  \label{Esp1E}
^{\circ\!}E_{ij }^{\alpha\beta} (k_\perp)=  \int\!  E _{ij}^{\alpha\beta}(k_\|, k_\perp)    d  k_\|      .~~
\end{eqnarray}
 Alternatively, one can average the 3D function  $\C E _{ij}(k_\|, k_\perp) $ over a plane. Here we choose  two planes:
\vskip  0.2 cm
 --  1D-spectra $ ^{\perp \!} E_{ij}^{\alpha\beta}(k_\|)$, averaged over a $\perp$\textit{-plane},  orthogonal  to $ \B U\sb{ns}$
 \begin{equation}  \label{Esp1C}
^{\perp\! } E_{ij}^{\alpha\beta} (k_\|)= \int\!   E _{ij}^{\alpha\beta}(k_\|, k_\perp )   dk_\perp  \ .
\end{equation}
 These spectra depend   on the streamwise projection $k_\|$  of the wave-vector $\B k$.

 --  1D-spectra  $ ^{\| \!} E^{\alpha\beta}_{ij}(k_\perp)$, averaged over the $\|$\textit{-plane},  oriented along  $ \B U\sb{ns}$. We chose for concreteness the plane $(k_x,k_z)$, such that $k_\|=k_x$ and  the spectra depend on $k_y$:
\begin{equation}  \label{Esp1D}
^{\|\! } E_{ij}^{\alpha\beta} (k_y)=\int  \C E _{ij}^{\alpha\beta}\Big (k_\|, \sqrt{k_z^2+k_y^2}  \ \Big ) \frac{dk_{||}\, dk_z}{4\pi^2}\ .
\end{equation}
 \end{subequations}

Note that the  2D-, and 1D-energy spectra are defined such that the  kinetic energy  density  per unit mass  $\C E_j$ can be found as:
 \begin{eqnarray}  \nn
2   E_{ j} &=&
   \int \C E_{jj}(\bm k)\, \frac{d^3 k}{(2\pi)^3}= \int \!\!\! E  _{ jj} (k_\|,k_\perp) d k_\|\, dk_\perp ~~~~~~ \\  \label{Esp2B}
   &=&\int  ~ \!\!\! \tilde E^{\alpha\beta}   _{j j} (k ,\theta)\, k\,  dk\,  d \cos\theta  =\!\int\! ^{\bullet\!} E_{jj}^{\alpha\beta} (k )  d k \\ \nn
     &=& \int \! ^{\perp\!} E_{jj }^{\alpha\beta} (k_\| )  d k_\|  = \! \int \! ^{\|\!} E_{jj }^{\alpha\beta} (k_y )  d k_y \ .
\end{eqnarray}
The tensor structure of the energy spectra will be considered only in \Sec{ss:tens}.   In the rest of
\Sec{s:DNS} we will restrict ourselves by  discussing  only  scalar versions of the energy spectra, which are the traces of their tensorial counterparts.

\begin{table*}[t]
\caption{\label{t:1}  Parameters of simulations by columns:
	(\#\,1) Run \#, (\# 2); Temperature  (K); 	(\# 3 and \# 4) Ratios of the  superfluid and normal-fluid  densities\cite{DB98}, $\rho\sb s / \rho \sb n$ and viscosities $\nu\sb s / \nu \sb n$ ;(\# 5) The mutual friction parameter $\alpha $;
	(\# 6 and \#7)  The numerical values of the kinematic viscosity of the normal-fluid and superfluid components  $\tilde \nu\sb n $  and   $\tilde \nu\sb s $;
	(\# 8 and \#9) The numerical values of mutual friction frequency $\Omega$ and counterflow velocity $V$;	(\# 10 and \#11)  the root mean square (rms)  normal-fluid and superfluid  turbulent velocity fluctuations $v\sp n\Sb T$ and $v\sp s\Sb T$;	(\# 12 and \# 13) the normal-fluid and superfluid Reynolds numbers;
	(\#~14)  $k_\times$,  \Eq{9}.\\
	For details see Sect.\ref{ss:procedure}. In all simulations: the number of collocation points along each axis is $N=256$;
	the computational box size  is $L=2\pi$,  the range of forced wavenumbers $k^{\tilde \varphi}=[0.5 , 1.5]$.  }
\begin{tabular*}{\linewidth}{@{\extracolsep{\fill} }    c    c    cc   c   c    c    c  c     c  c c c    c}
	\hline\hline
	1   & 2      & 3                &4  & 5 & 6                         & 7                                     & 8           & 9          & 10                             & 11                             & 12         & 13         & 14         \\ \hline
	Run & $T$,   & $\displaystyle \frac{\rho\sb s }{ \rho \sb n}$ & $\displaystyle \frac{\nu\sb s }{ \nu \sb n}$ &  ~~$\alpha$~~&$\tilde \nu\sb n$ , & $\tilde \nu \sb s $,  & $\Omega$, & ~$V~$& $u\sp n \Sb T=$               & $u\sp s \Sb T=$               & Re$\sb n $ & Re$\sb s $ & $k_\times$ \\
	\#  & K      &                                                && & $\times 10^3$             & $\times 10^3$              &           &            & $ \sqrt {\< {u\sb n}^2 \> } $ & $ \sqrt {\< {u\sb s}^2 \> } $ &            &            & $  ~ $     \\ \hline
	1   &        &                                                & &                           &                            &              & 1          & 15         & 4.3                           & 4.5                           &   997         &  2595         & 0.34       \\
	2   & 1.65~ & 4.2                                           &0.49   & 0.11& 3.0                      & 1.38                             & 20         & 15         & 4.2                           & 4.2                           &  1216         &    2667       & 6.88       \\
	3   &        &                                                & &                           &                            &              & 1          & 0          & 3.6                           & 3.6                           &  1257          & 2953           & $\infty$   \\
	4   &        &                                                & &                           &                            &              & 20         & 0          & 3.7                           & 3.7                           &  1316          &2872            & $\infty$   \\ \hline
	5   &        &                                                & &                           &                            &              & 1          & 15       & 3.4                           & 3.5                           &  908          &    994        & 0.18       \\
	6   & 1.85~ & 1.75                                          &1.07   & 0.18 & 3.0                      & 1.85                            & 20         & 15      & 3.5                             & 3.5                             &                  1051                         &   1056              & 3.57       \\
	7   &        &                                                & &                           &                            &              & 1          & 0          & 3.3                           & 3.3                           &  1154          &   1239         & $\infty$   \\
	8   &        &                                                & &                           &                            &              & 20         & 0          & 3.6                           & 3.5                           &    1179       &    1181        & $\infty$   \\ \hline
	9   &        &                                                & &                           &                            &              & 1          & 15         & 4.3                           & 4.2                           &     1064      &     582       & 0.15                               \\
	10  & 2.00~   & 0.83                                          & 1.72 & 0.28& 3.0                   & 5.0                                   & 20         & 15         & 3.5                           & 3.5                                      &    1153         &   664       & 1.5                   \\
	11  &        &                                                & &                           &                            &              & 1          & 0          & 3.3                           & 3.3                           &  1225          &  689         & $\infty$   \\
	12  &        &                                                & &                           &                            &              & 20         & 0          & 3.6                           & 3.5                           &    1177         &  676          & $\infty$   \\ \hline\hline
	&        &                                                 & &                           &
\end{tabular*}

\end{table*}
\begin{figure*}
\begin{tabular}{ccc}
	(a)   &(b)  &(c)  \\
	\includegraphics[scale=0.29]{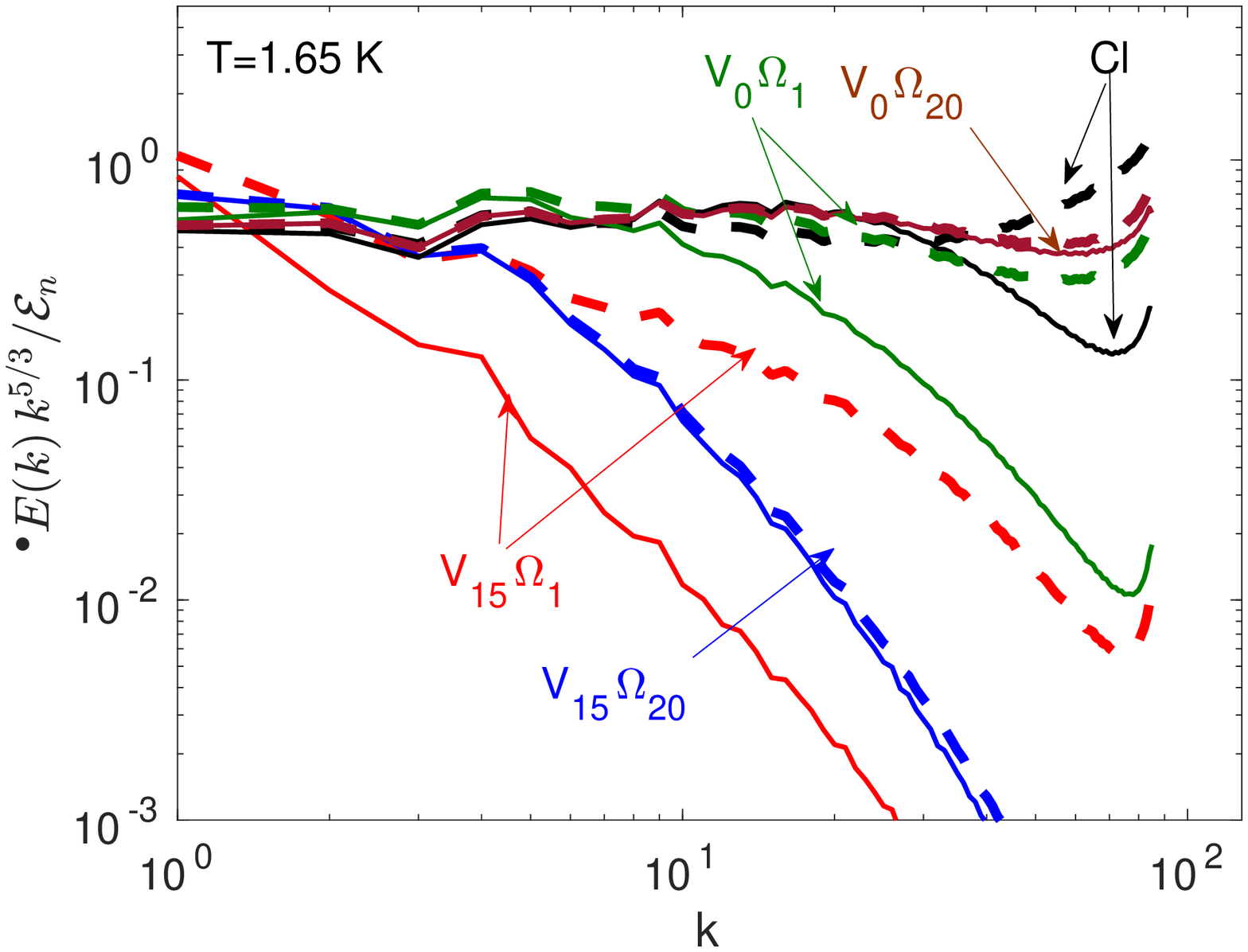}&
	\includegraphics[scale=0.29]{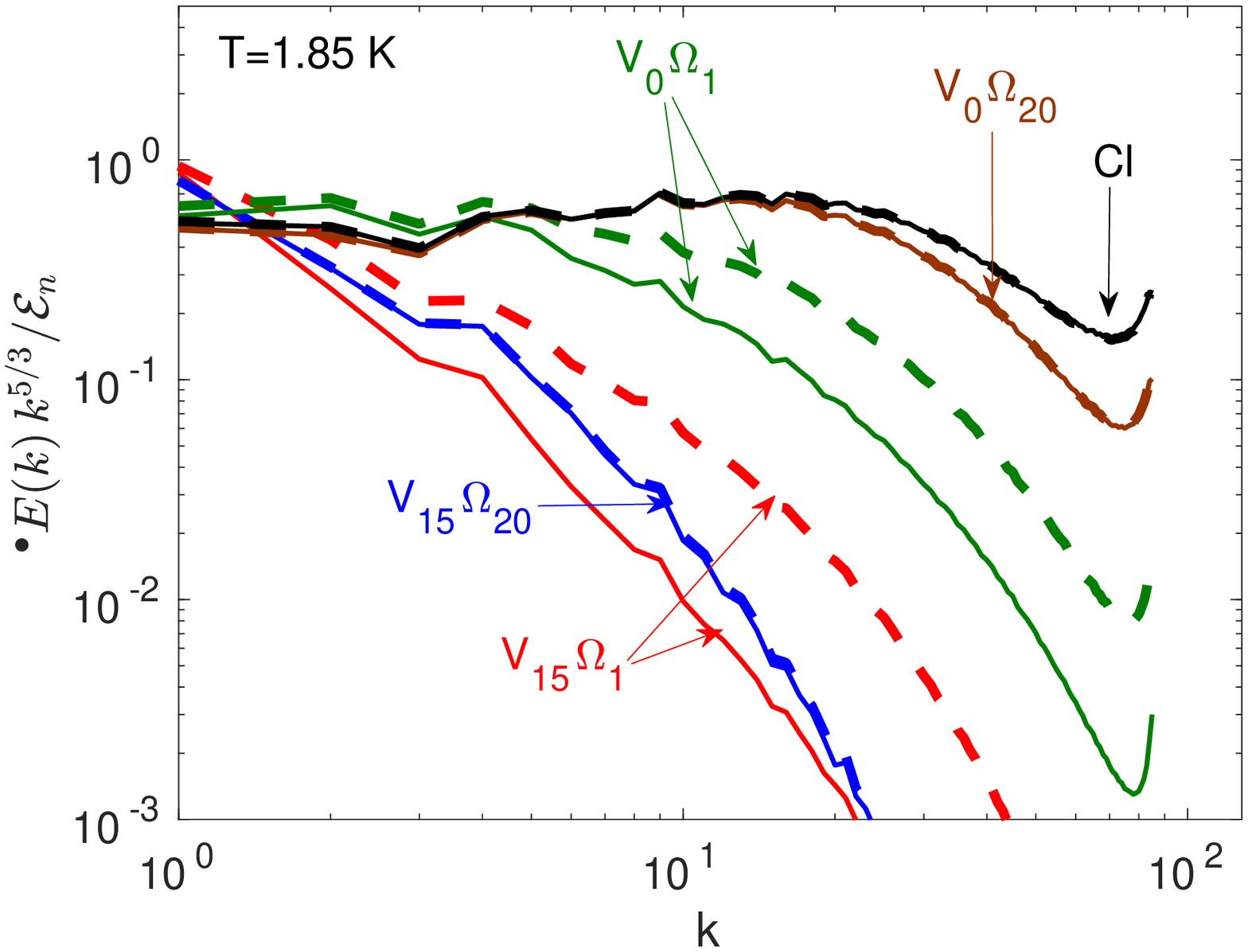}&
	\includegraphics[scale=0.29]{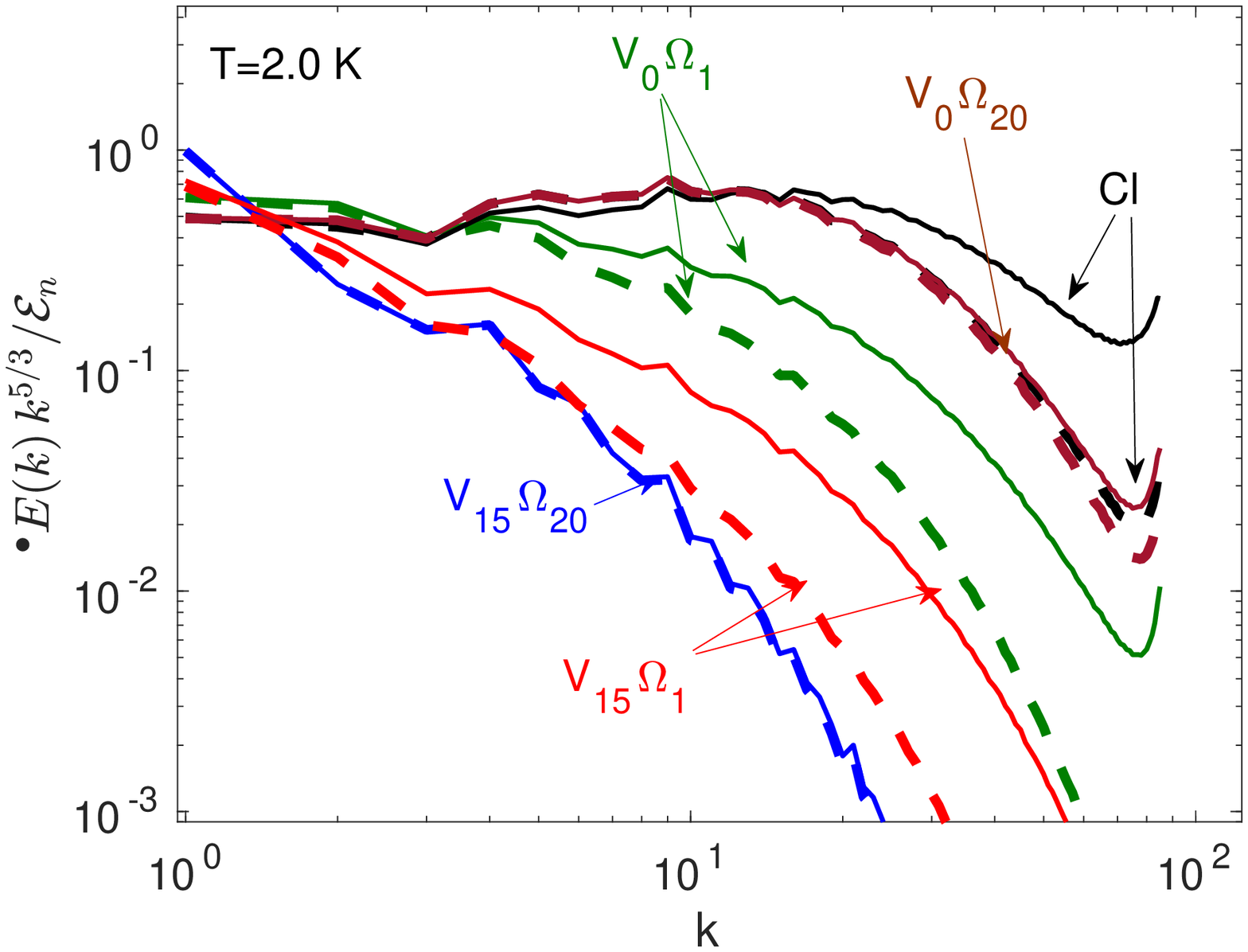}\\ 		
	(d)   &(e)  &(f)\\
	\includegraphics[scale=0.29]{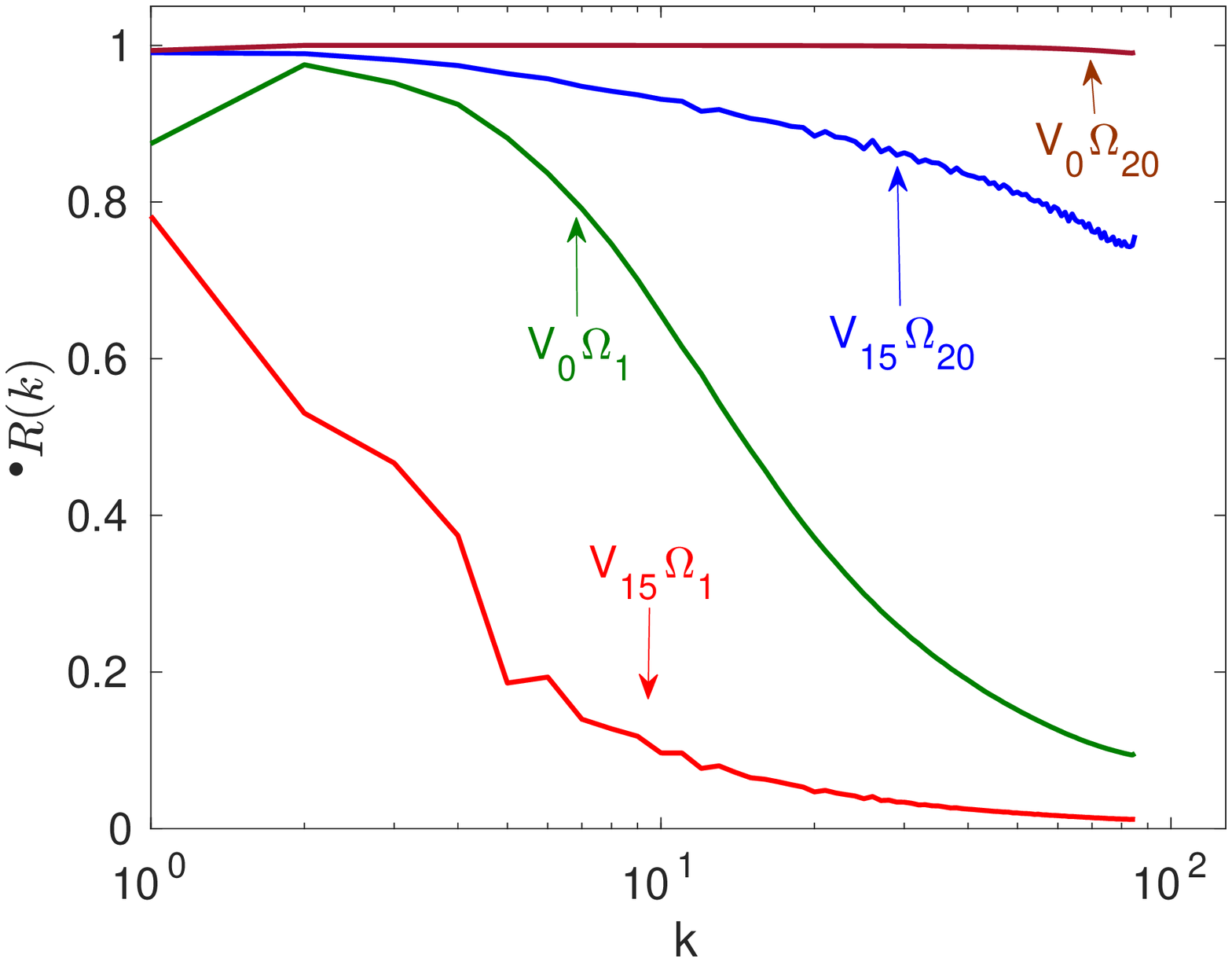}&
	\includegraphics[scale=0.29]{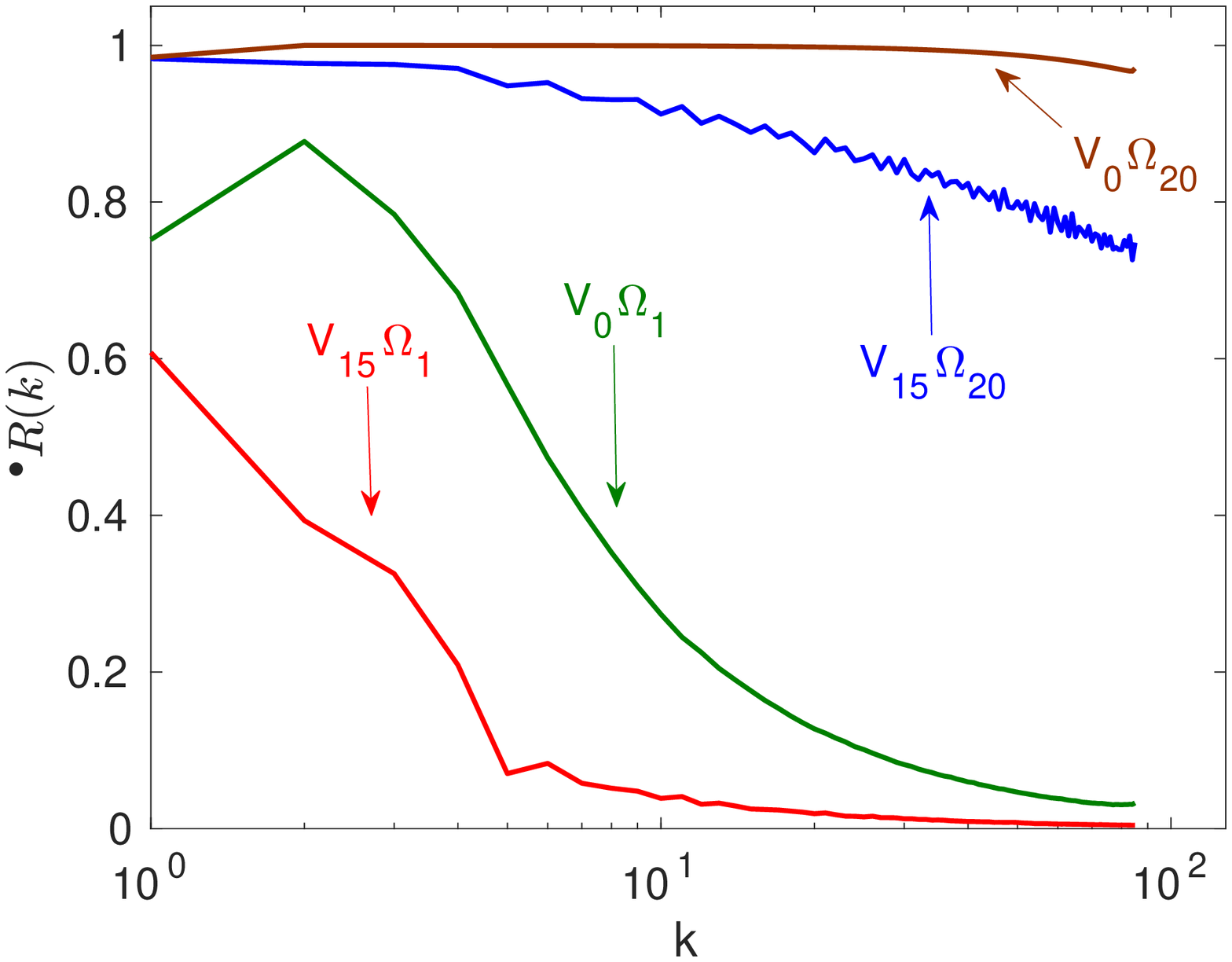}&
	\includegraphics[scale=0.29]{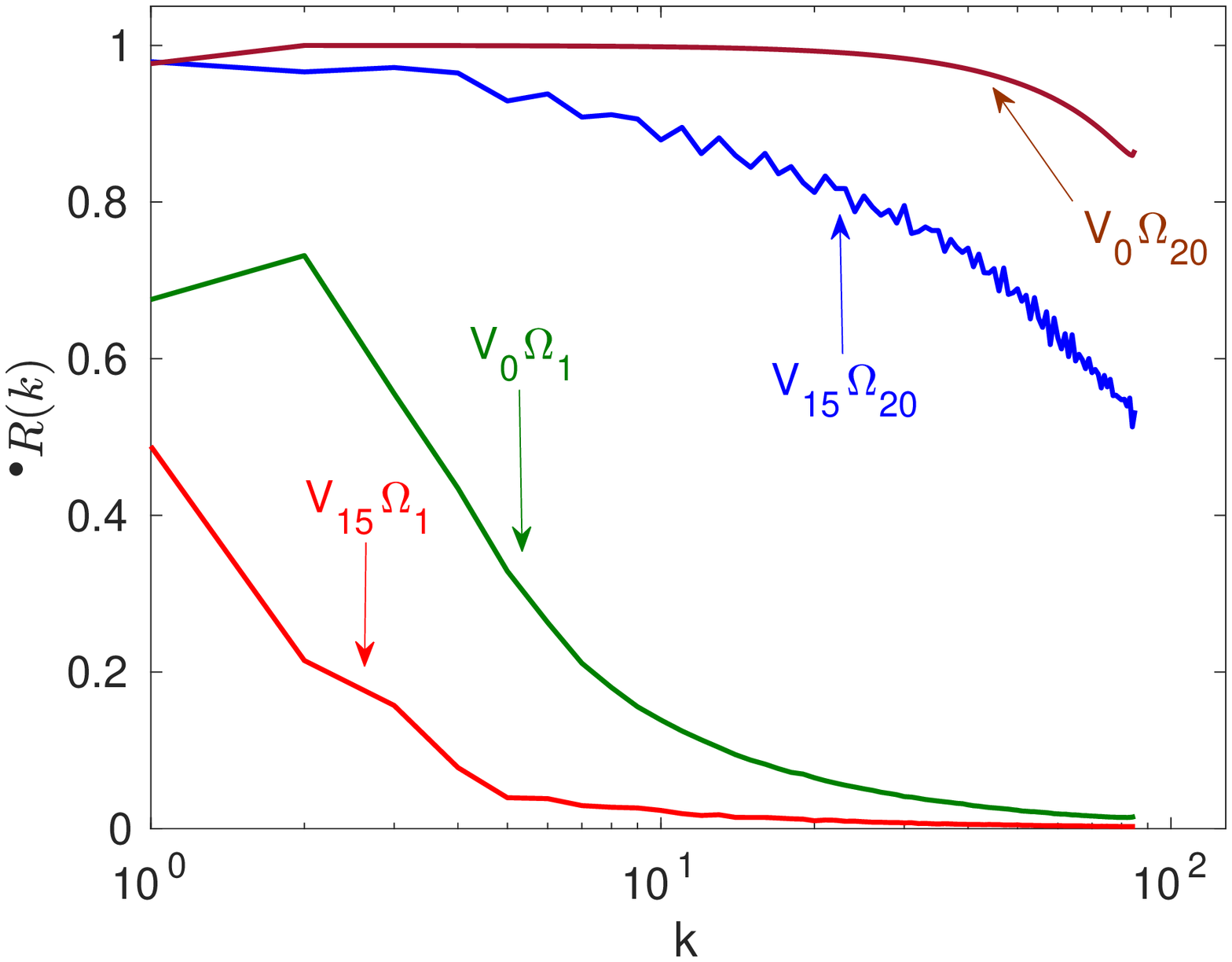}\\
\end{tabular}
\caption{\label{f:1} The spherically-averaged energy spectra and the cross-correlation functions for the counterflow and coflow at different temperatures.  Panels (a),(b) and (c) -- the K41-compensated energy spectra  $^{\bullet\!}E_j(k)$ for the normal-fluid (solid lines) and superfluid components (dashed lines).  Panels (d),(e) and (f) -- the normalized cross-correlation functions $^{\bullet\!}R(k)$. The four sets of lines in each panel correspond to $V_0\Omega_1$-- green lines, $V_0\Omega_{20}$-- brown lines, $V_{15}\Omega_{1}$-- red lines, $V_{15}\Omega_{20}$-- blue  lines. The black lines, labeled "Cl"  in the  panels (a),(b) and (c) correspond to the spectra of  classical turbulence. }

\end{figure*}
 \subsubsection{\label{ss:VSF}Velocity structure functions}
Another presentation of the statistics of turbulence is provided by the second-order velocity structure functions
 	\begin{subequations}
 		\begin{eqnarray}\label{S2a}
 		\delta_{_{\B R}}u^\alpha _j&\=& u^\alpha _j(\B R+\B r,t)-u^\alpha _j(\ \B r,t)	\\ \label{S2b}
 		S^{\alpha\beta}_j(\B R)&\=&\< \delta_{_{\B R}}u^\alpha _j \, \delta_{_{\B R}}u^\beta _j   \>\ .
 		\end{eqnarray}\end{subequations}
 The trace $S_j(\B R)\= \sum_\alpha   S_j^{\alpha\alpha}(\B R)$ is
a measure of the kinetic energy of turbulent (normal or superfluid) velocity fluctuations on scale $R$.
Recently,  the streamwise normal velocity across a channel,
 	$v^x(y,t)$  was measured using thin lines of the
 	triplet-state He$_2$ molecular tracers created by a femptosecond-laser
 	field ionization of He atoms\,\cite{18} across the channel.  This way, the transversal 2$\sp {nd}$-order structure functions\,\cite{WG-2015,WG-2017} of the normal-fluid velocity differences $S\sb n^{xx} (R_y)$ were obtained. Similarly, one can use two  or more tracer lines, separated in the stream-line direction $\B {\^ x}$, to measure the longitudinal structure function  $S\sb n^{xx} (R_x)$ and even inclined structure function $S\sb n^{xx} (R_x,R_y)$.

Using the definition of the structure functions\,\eqref{S2a} and the one-dimensional version of the inverse Fourier transform	\,\eqref{def-Fy} one gets
 	\begin{subequations}\label{S2-1}
 		\begin{eqnarray}\label{S2-1a}
 		S_j^{xx} (R_y) &=& 8 \int\limits  _0^\infty {^\parallel\!  E}_j^{xx}(k_y) \sin^2 \frac {k_y R_y}{2}\,  dk_y\,,	\\ \label{S2-1b}
 		S_j^{xx} (R_x) &=&  8 \int\limits  _0^\infty {^\perp\!  E}^{xx}\sb n (k_x)\sin^2 \frac {k_x R_x}{2}\,  dk_x	\ . 		
 		\end{eqnarray}
 	\end{subequations}
 	Analyzing the integrals\,\eqref{S2-1} for the scale-invariant spectra $E(k)\propto k^{-m}$ one concludes that they converge in the \textit{window of locality}
 	\begin{subequations}\label{S2-2}
 		\begin{equation}\label{S2-2a}
 		1< m < 3 \ . 		
 		\end{equation}
 		In this window, the leading contribution to the integrals\,\eqref{S2-1}
 		comes from the region $k R \sim 1$ and
 		\begin{equation}\label{S2-2b}
 		S_j(R) \propto R^n\,, \quad n=m-1	\ .
 		\end{equation}
 		This is a well known relationship. For example, $n = 2/3$  for the K41
 		spectrum with $m = 5/3$ [which satisfy \eqref{S2-2a}]. However for fast decaying spectra with $m \geq 3 $
 		the integrals\,\eqref{S2-1a} diverge in the infrared region $k R \ll 1$ with the main contribution coming from energy containing region $k\sim k_0$, giving
 		\begin{equation}\label{S2-2c}
 	S^{\alpha \alpha}_j(R) \propto R^2 	\ .
 		\end{equation}
 	\end{subequations} 
 	We see that connection between $m$ and $n$ for fastly decaying spectra with $m>3$ is lost. To recover it for  $m>3$ we consider structure functions of the velocity second differences\cite{Luca1,Luca2}
 		\begin{subequations}
 			  \begin{eqnarray}\label{S3a}
 			&&\hskip -2 cm 	\D_{_{\B R}}u^\alpha _j \=  u^\alpha _j(\B 2 R+\B r,t)- 2 u^\alpha _j( \B R + \B r,t)	+ u^\alpha _j(\ \B r,t),~~~~~\\ \label{S3b}
 			\~ S^{\alpha\beta}_j(\B R)&\=&\< \Delta_{_{\B R}}u^\alpha _j   \Delta_{_{\B R}}u^\beta _j   \>\ .
 			\end{eqnarray}
 		\end{subequations}
 		Now instead of \Eqs{S2-1} we have
 		\begin{subequations}\label{TS3-1}
 			\begin{eqnarray}\label{TS3-1a}
 			\~S_j^{xx} (R_y) &=& 32 \int\limits  _0^\infty {^\parallel\!  E_j}^{xx}(k_y) \sin^4 \frac {k_y R_y}{2}\,  dk_y\,,	\\ \label{TS3-1b}
 			\~S_ j^{xx} (R_x) &=&  32 \int\limits  _0^\infty {^\perp\!  E}_j^{xx} (k_x)\sin^4 \frac {k_x R_x}{2}\,  dk_x	\ . 		
 			\end{eqnarray}
 		\end{subequations}
 		Now these integrals converge  in the extended \textit{window of locality}
 		\begin{subequations}\label{S3-2}
 			\begin{equation}\label{S3-2a}
 			1< m < 5 \ . 		
 			\end{equation}
 			In this window, the leading contribution to the integrals\,\eqref{TS3-1}
 			again comes from the region $k R \sim 1$ and, similarly to \Eq{S2-2b}, for the scale-invariant spectrum $E(k)\propto k^{-m}$ we have
 			\begin{equation}\label{S3-2b}
 			\~ S^{\alpha \alpha}_j(R) \propto R^n\,, \quad n=m-1	\ .
 			\end{equation}
 			For $m>5$ the integrals\,\eqref{TS3-1} diverge in the infrared region
 			and
 			\begin{equation}\label{S3-2c}
 		\~ S^{\alpha \alpha}_j(R) \propto R^4\  .
 			\end{equation}
 		\end{subequations}	

 It is worth noting that the relations \eqref{S2-2b} and \eqref{S3-2b} are valid in the limit of infinite inertial interval. For a finite inertial interval which is typical for the experimental conditions, the structure functions have a complicated functional dependence, mixing the inertial and viscous behavior and the original scaling of the energy spectra is reproduced over very short intervals of scales\cite{WG-2018}.
 Nevertheless, when experimental conditions do not allow to measure the energy spectra directly, the structure functions remain the preferred tool to access the statistics of the velocity fluctuations. In a turbulent  counterflow, where the energy spectra are not scale-invariant, the  quantitative analysis of the structure functions may not be meaningful.  Nevertheless a qualitative difference between structure functions measured along different directions may confirm the presence of the spectral anisotropy.
 	
 	\begin{figure*}
 		\begin{tabular}{ccc}
 		\end{tabular}
 		\begin{tabular}{ccc}
 			(a)   & (b)  & (c)  \\	
 			\includegraphics[scale=0.31]{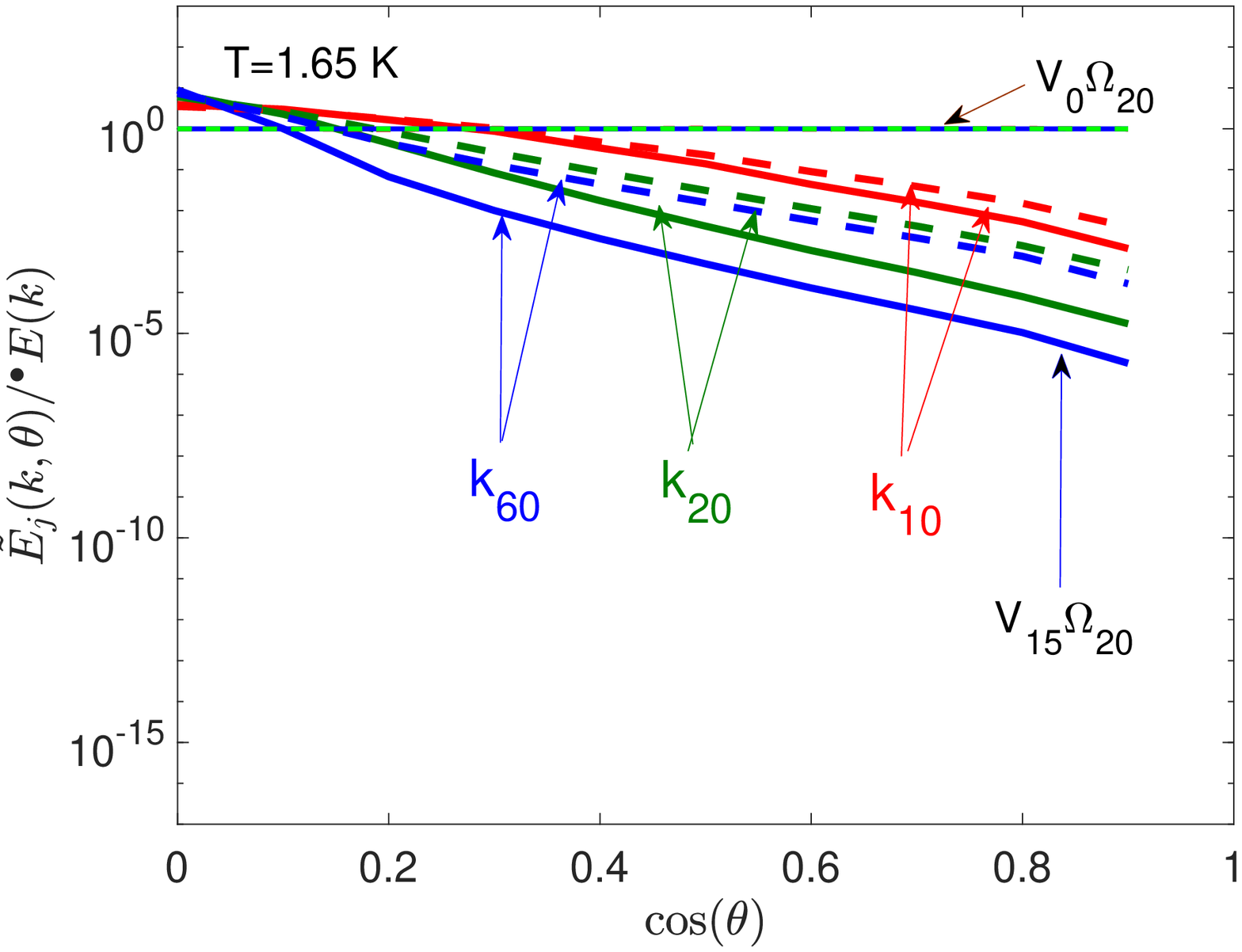}&
 			\includegraphics[scale=0.31]{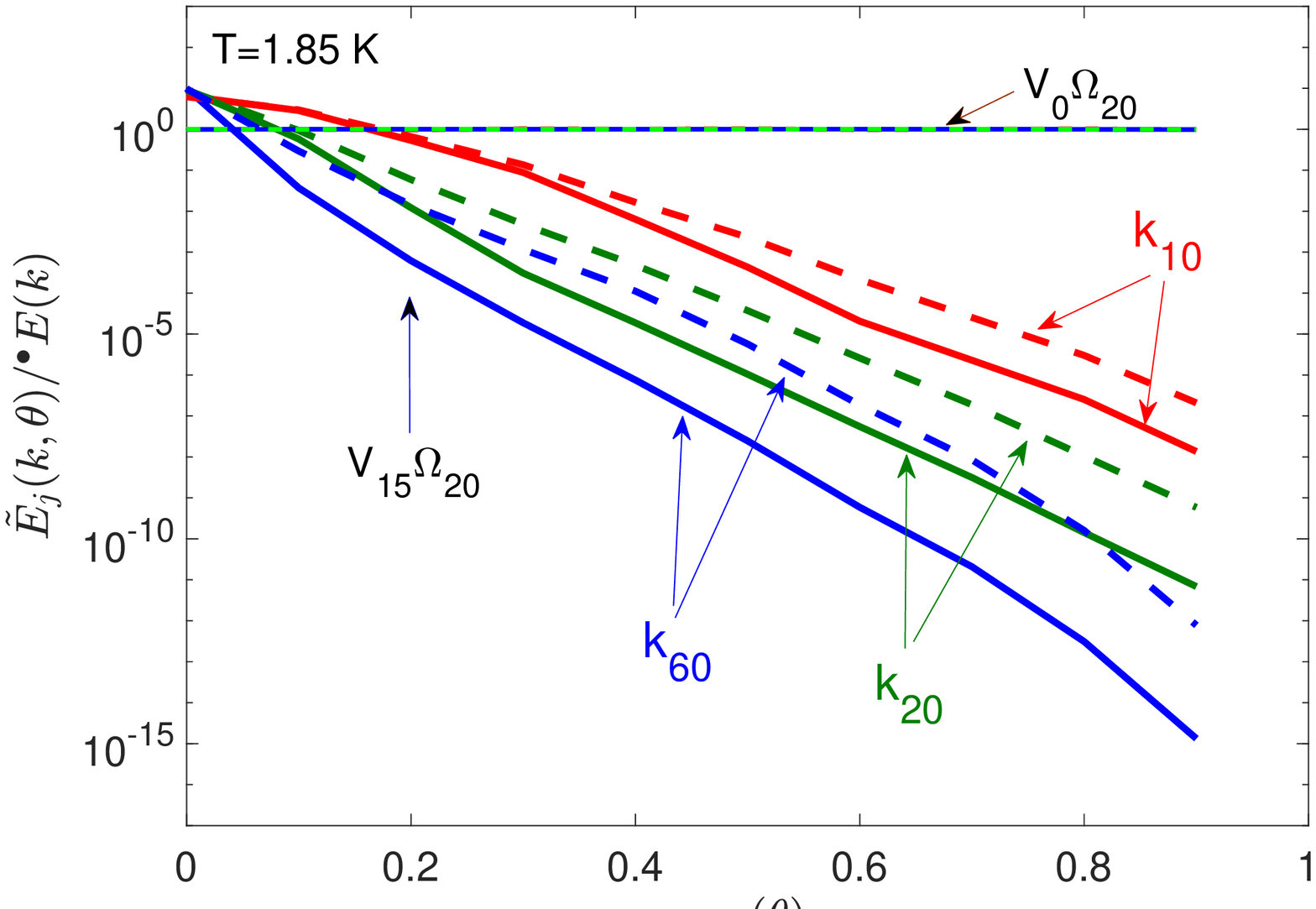}&
 			\includegraphics[scale=0.31]{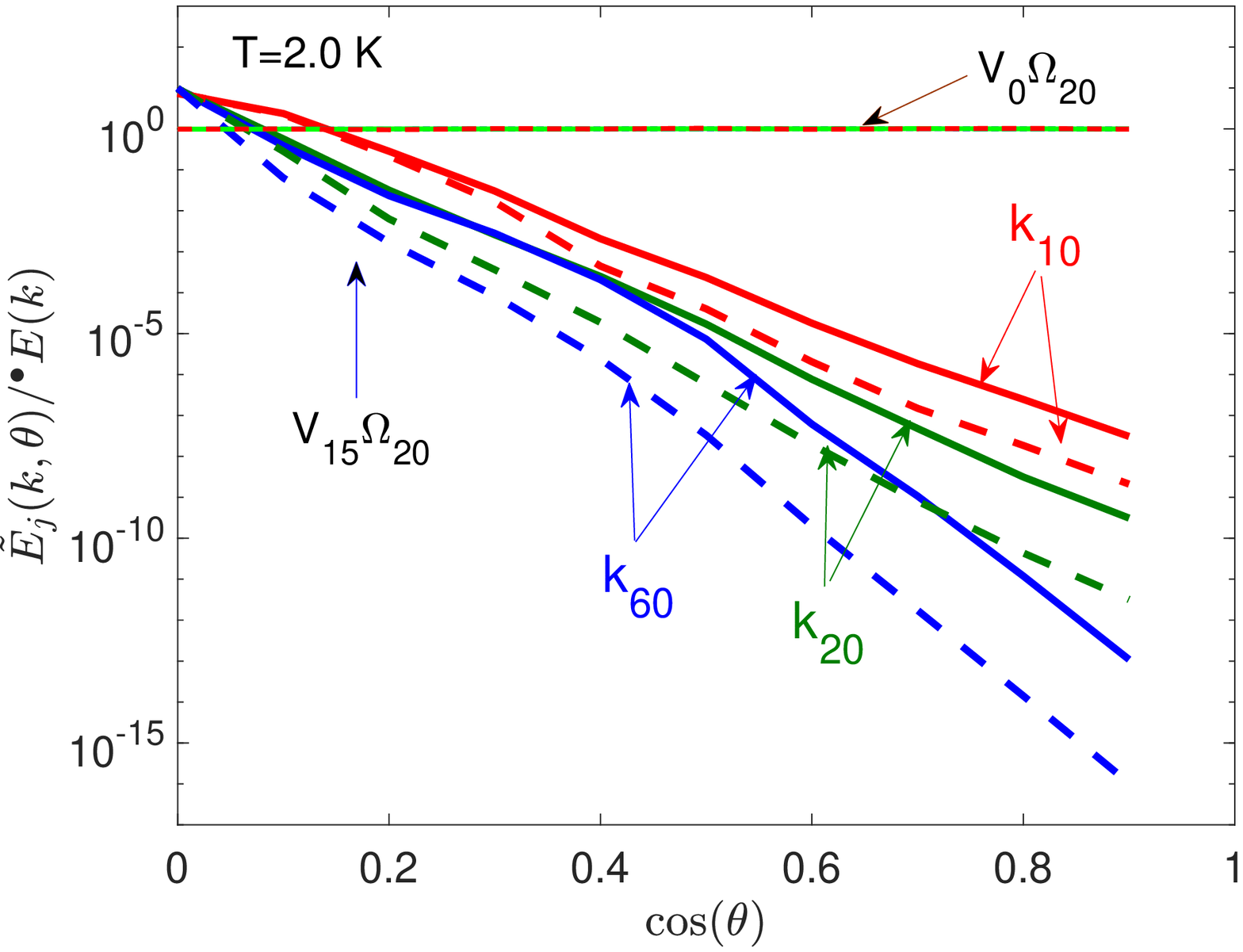}\\
 			(d) & (e)  & (f) \\	
 			\includegraphics[scale=0.31]{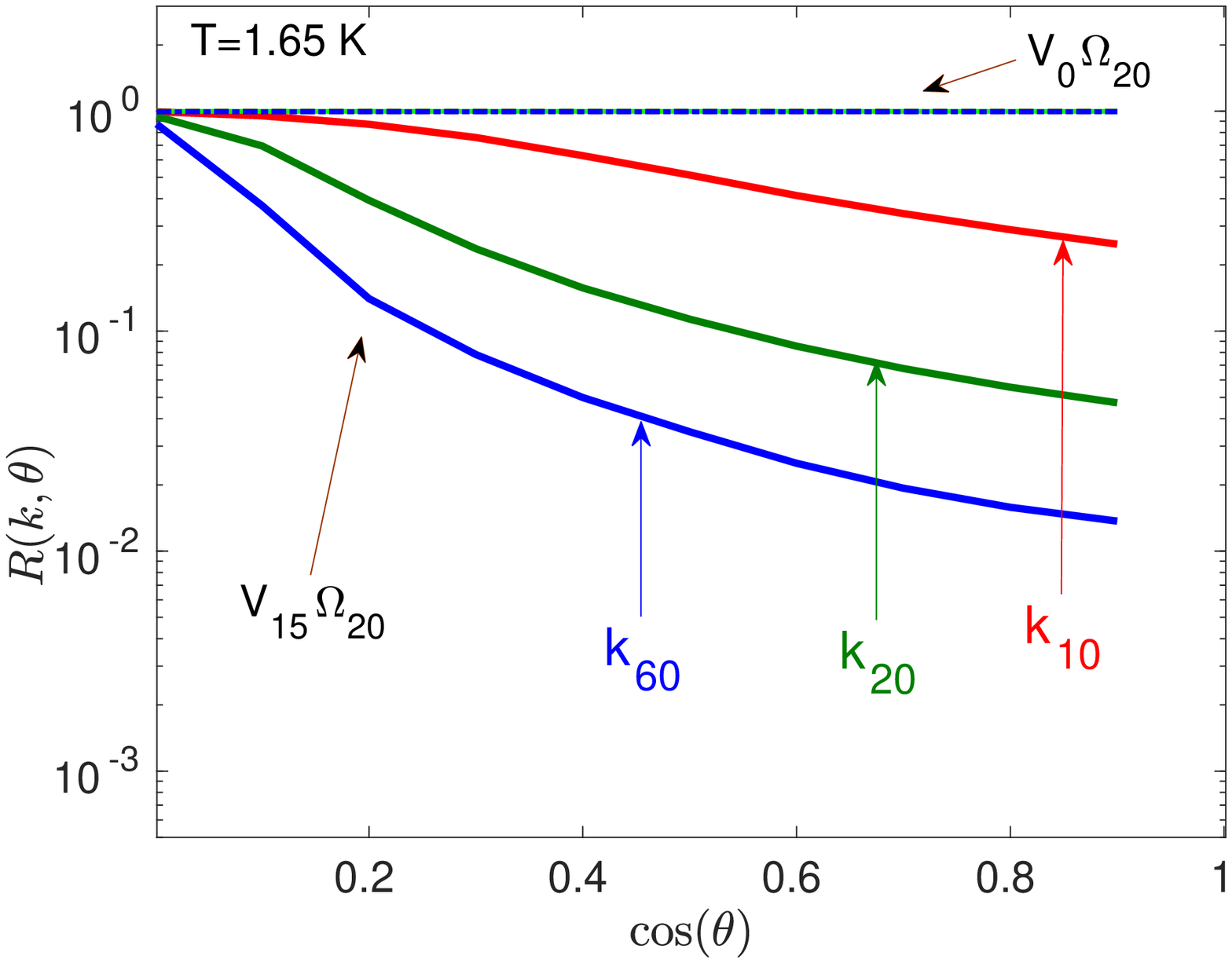}&
 			\includegraphics[scale=0.31]{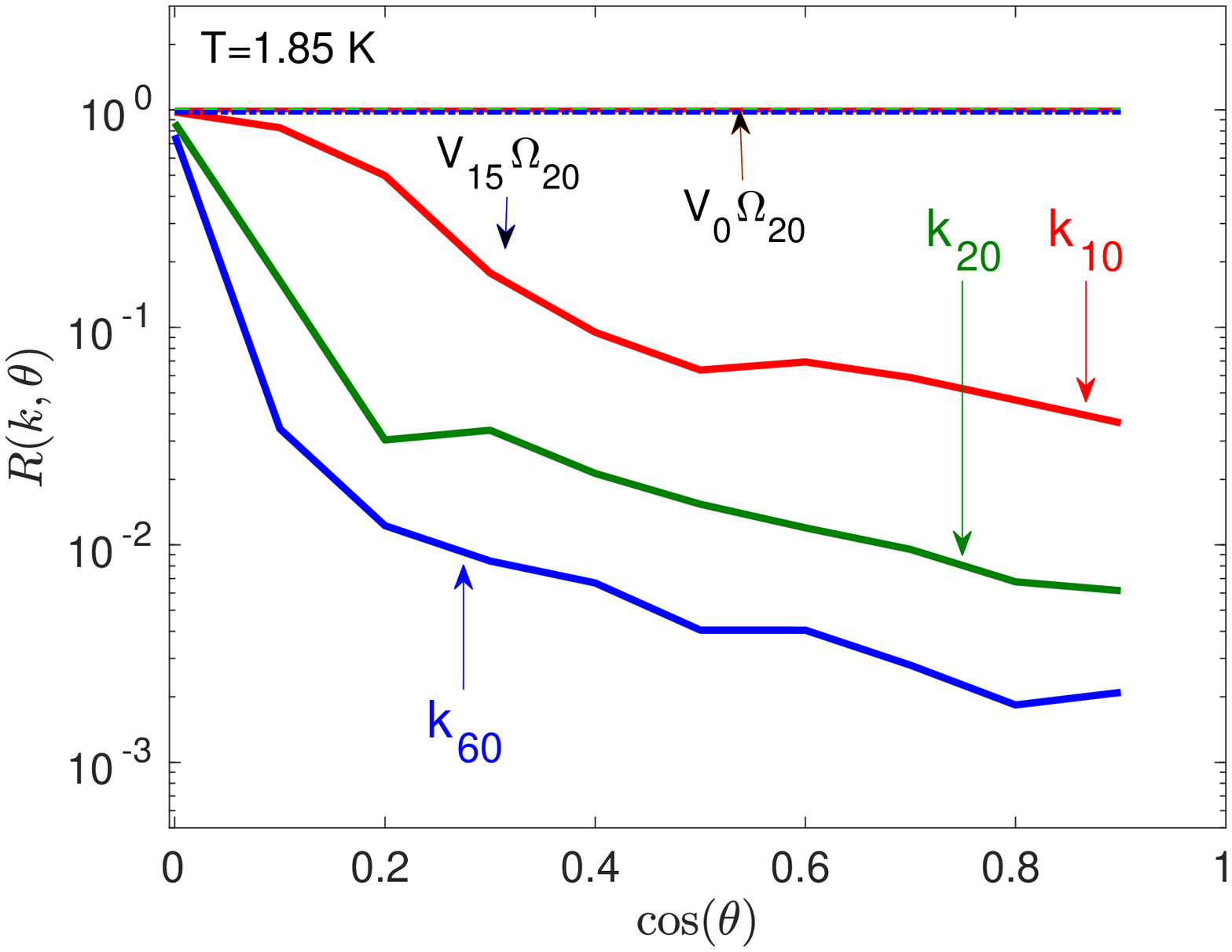}&
 			\includegraphics[scale=0.31]{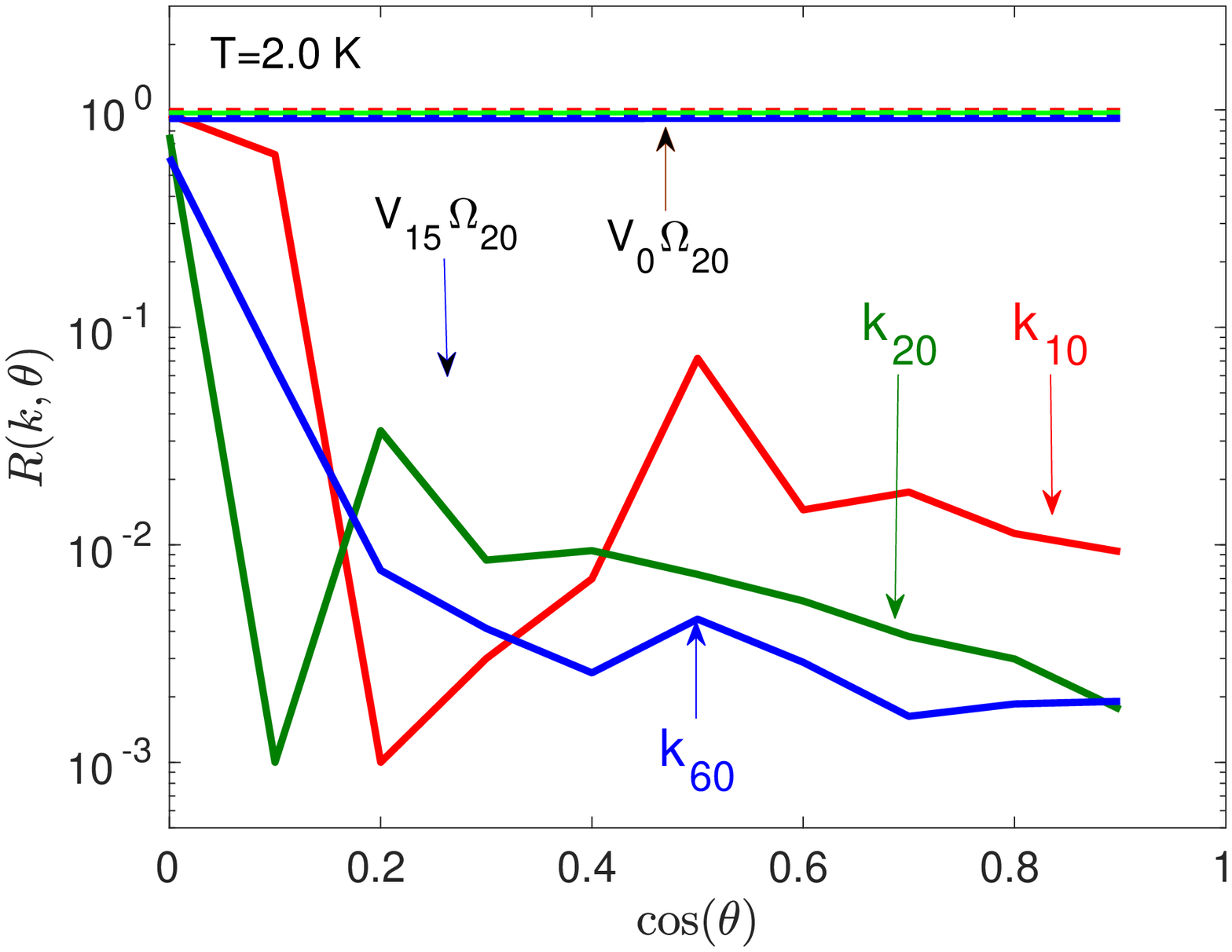}\\
 			
 		\end{tabular}
 		\caption{ \label{f:2}  The angular dependencies of the  energy spectra $\tilde E_j(k,\theta)/\, ^{\bullet\!}E_j(k)$ and the cross-correlations $R (k,\theta)$, averaged in 3 wavenumber bands, for various flow condition. Panels (a),(b) and (c)- the 2D energy spectra  $\tilde E_j(k,\theta)$. The spectra of the normal-fluid  are shown by solid lines and of the superfluid -- by dashed lines. Panels (d)-(f)-- the cross-correlations  $R (k,\theta)$.
 			The spectra and the cross-correlation  for the coflow  are shown by thin horizontal lines and marked $V_0\Omega_{20}$,  for the counterflow -- by thick  lines and marked $V_{15}\Omega_{20}$.  In all panels, red lines correspond to the averaging over wavenumber range $10\leq k< 20$ (labeled as $k_{10}$), green lines -- to averaging over $20\leq k< 60$  (labeled as $k_{20}$) and blue lines -- to the averaging over $60\leq k\le 80$  (labeled as $k_{60}$).   Note the log-linear scale. }
 		
 	\end{figure*}

\subsection{\label{ss:theory}Physical origin of the strong anisotropy of counterflow turbulence}

In a recent Letter\cite{AnisoLetter} it was shown that the energy spectra in counterflow turbulence are expected to be strongly anisotropic.  To keep the present paper self-contained, we repeat here some of that discussion and add further clarifications to the analysis.

We start with a balance equation\cite{LP-2018,AnisoLetter} for the 2D energy spectrum   $\tilde E_j(k,\theta)$ in the  counterflow turbulence with axial symmetry:
\begin{subequations}\label{balance1}
\begin{eqnarray}\label{balance1A} \frac{\partial \tilde E_j(k,\theta,t)}{\partial t} \!\!&\!\!+\!\!&\!\!
\mbox{div}_{\B k } [\B \ve_j ( \B k)]  = - \C D_j\sp{mf}(k,\theta ) - \C D_j^{\nu}(k,\theta ) ,  ~~~~~~\\
\label{balance1B}
\C D_j\sp{mf}(k,\theta )&=&\Omega_j [ \tilde E_j (k,\theta)- \tilde E\sb{ns}(k,\theta)  \big ] \,, \\  \label{balance1C}
\C D_j^{\nu}(k,\theta )&=& 2\, \nu_j k^2\tilde E_j(k,\theta)\ .
\end{eqnarray}
\end{subequations}
Here  $\mbox{div}_{\B k } [\B \ve_j ( \B k)] $ is the transfer term due to inertial nonlinear effects.
The terms on the right hand side  describe the energy dissipation rate due to  the mutual friction  $\C D_j\sp{mf}(k,\theta )$ and due to the viscous effects $\C D_j^{\nu}(k,\theta )$. To keep the presentation  concise, we introduced a notation  $\Omega\sb n = \Omega\sb s \rho\sb s/ \rho \sb n$.

The origin of the  energy spectra anisotropy in counterflow turbulence can be deduced from the form of  the dissipation rate  $\C D_j\sp{mf}(k,\theta )$   \eqref{balance1B}.
In this term, the cross-correlation function $\tilde E \sb{ns}(k,\theta)$  has the following form [cf.  Eq.(13) in \Ref{decoupling}]:
\begin{subequations}\label{LP-20} 
	\begin{equation}\label{LP-20A}
\tilde E\sb{ns}(k,\theta)= \frac{AB}{B^2+ (\B k\cdot \B U\sb{ns})^2}\ .
\end{equation}
Here $A= \Omega\sb s \tilde E\sb n (k,\theta)+ \Omega\sb n \tilde E\sb s (k,\theta)$  and $B$  can be written\cite{LP-2018} as $B=\Omega\sb{ns}=\Omega\sb n + \Omega \sb s$.  We further note \cite{LP-2018,AnisoLetter}that when two
components are highly correlated, the cross-correlation
may be accurately represented by the corresponding energy
spectra. For wave numbers where the components are not
correlated, $\tilde E\sb{ns}(k,\theta)$  is small and the accuracy of its
representation is less important.
This allows us to decouple $\tilde E\sb{ns}(k,\theta)$ in \Eqs{balance1} for  each component as follows:
\begin{eqnarray}\label{LP-20B}
\tilde E\sb{ns}(k,\theta)&=&\tilde E_j(k,\theta) D(k,\theta)\,,  \\ \label{LP-20C}
D(k,\theta) &=&1\Big / \Big [1+ \Big (\frac{k U\sb {ns}\cos \theta}{\Omega\sb{ns}}\Big )^2 \Big]  \ .
\end{eqnarray}
Note that averaging  \Eq{LP-20C} over $\theta$ results in the equation for $D(k)$,
used in the theory of isotropic counterflow turbulence\,\cite{LP-2018}:
\begin{eqnarray}  \nn
D(k)&\=&\< D(k,\theta)\> _\theta=  \int\limits_0^1  D(k,\theta) d \cos \theta \\ \label{9}
& = &\frac{k_\times}{k }\arctan \frac{k  }{k_\times}\,, \quad k_\times\= \frac{\Omega\sb{ns}}{U\sb{ns}}\ .
\end{eqnarray}
\end{subequations}

The function $D(k,\theta)$ in \Eqs{LP-20} describes the level of decorrelation of the  normal-fluid and superfluid velocity components by the counterflow velocity. Within the approximation  \eqref{LP-20B},   $D(k,\theta)$ defines the rate of energy dissipation caused by mutual friction:
\begin{equation}\label{diss}
\C D_j\sp{mf}(k,\theta)= \Omega_j \tilde E_j\big [ 1- D(k,\theta) \big]\ .
\end{equation}
 For small $k$ or even for large $k$ with $\B k$  almost perpendicular  to  $\B U\sb{ns}$ (i.e $\cos\theta\ll 1$), $D(k,\theta)\simeq 1$ , the normal and superfluid velocities are almost fully coupled and the rate of the energy dissipation $\C D_j\sp{mf}(k,\theta)\ll \Omega_j$ is small.  In this case, the role of mutual friction is minor  and we  expect the energy spectrum $\tilde E_j(k,\theta)$ to be close to the classical prediction
$E\Sb{K41}(k)\propto k^{-5/3}$.   For large $k$ and for $\B k$ with $\cos\theta\sim 1$,  $D(k,\theta) \ll  1$, the  velocity components are almost decoupled, $\C D_j(k,\theta)\ll 1$ and the mutual-friction energy dissipation is maximal: $\C D_j\sp{mf}(k,\theta)\approx \Omega_j$. This situation is similar to that in $^3$He with the normal-fluid component at rest, for which $\C D\sb s\sp{mf}(k,\theta)= \Omega\sb s$.   In this case, we can expect that the energy dissipation by mutual friction  strongly suppresses the energy spectra, much below the K41 expectation $E\Sb{K41}(k)$ up to the level typical for the $^3$He turbulence\,\cite{LNV,DNS-He3,He3,LP-2017}.
Next, we note that $\Omega\sb{ns}=\Omega\sb s+\Omega\sb n=\rho\, \Omega\sb s/\rho \sb n$. Then \Eq{LP-20C} may be rewritten as
	\begin{equation}
D(k,\theta) =1\Big / \Big [1+ \Big (\frac{\rho\sb n k U\sb {ns}\cos \theta}{\rho \Omega\sb{s}}\Big )^2 \Big] .
	\end{equation}
	At low temperatures $\rho\sb n/\rho$ is small and the velocity decorrelation is considerable only at large $k$, while at higher temperature  $\rho\sb n/\rho\approx 1$ and the energy dissipation by mutual friction is effective at all scales.

Combining all these considerations, we expect the energy spectra $\tilde E_j(k,\cos \theta)$  to become more  anisotropic with increasing $k$, with most of energy concentrated in the range of small $\cos \theta$, i.e. in the wavevector plane orthogonal to the counterflow velocity $\B U\sb{ns}$. This effect is milder at low $T$ and stronger at higher temperatures.

\section{\label{s:DNS}Strong anisotropy of energy spectra}

\subsection{\label{ss:procedure} Simulation parameters and numerical procedure}

The direct numerical simulation of the coupled \Eqs{NSEs} 
 were carried out using a fully de-aliased pseudospectral code with a resolution of $256^3$ collocation points in a triply
periodic domain of size $L=2\pi$. The parameters of the simulations are summarized in Table~\ref{t:1}.
To obtain the steady-state evolution,  velocity fields of the normal and superfluid components are stirred by two independent isotropic random Gaussian forcings:

\begin{equation}\label{force}
\langle {\tilde{\B  \varphi}}_u({\bm k},t)\cdot { \tilde{\B \varphi}}_u^*({\bm k}',t') \rangle =\Phi(k) \delta( {\bm k}-{\bm k}') \delta(t-t') \widehat P({\bm k})\,,
\end{equation}
where $\widehat P({\bm k})$ is a projector assuring incompressibility
and $\Phi(k)=\Phi_0 k^{-3}$;  the forcing amplitude $\Phi_0$ is nonzero only in a given band of Fourier modes:
$ k \in [0.5,1.5]$. Both components are forced with the same amplitude to allow direct comparison with simulations of the uncoupled equations. The time integration is performed using a 2-nd order Adams-Bashforth scheme with the viscous term exactly integrated.
Simulations for all  temperatures   were carried out with the  normal-fluid viscosity fixed at $\tilde \nu\sb n=0.003$ and
the value of  $\tilde \nu\sb s$ is found using the known value of ratio $\nu\sb s/\nu\sb n$ at each temperature.

To properly expose various aspects of the counterflow turbulence statistics we chose several sets of governing parameters for the simulations.
Since the material parameters of \He4 are strongly temperature dependent\cite{DB98} (see Table \ref{t:1}, columns  \#3--5), we consider three temperatures, corresponding to an experimentally accessible range $T=1.65, 1.85$ and $2.0$K. At low temperatures the superfluid component is dominant and has a lower viscosity, while at high $T$ the density of the normal-fluid component is larger, while its kinematic viscosity is lower. At $T=1.85$K, the densities and the viscosities of two components are closely matched.

As was shown in previous studies\cite{decoupling,LP-2018, BLPSV-2016}, the major role in the statistics of the counterflow in superfluid \He4 is played by the ratio of the mutual friction frequency and the counterflow Doppler frequency $\Omega\sb{ns}/(k U\sb{ns}$) [cf. \Eq{9}]. To explore various scenarios, we use one counterflow velocity $U\sb{ns}$ and two very different values of $\Omega\sb{s}$.  To emphasis the importance of the flow anisotropy, we compare the results for the counterflow with simulations of coflow $U\sb{ns}=0$,  keeping the rest of parameters unchanged. The detailed study of the statistics of the coflow was reported in \Ref{DNS-He4}. Other parameters of the simulations were chosen based on dimensionless numbers: i) the Reynolds numbers  Re$_j=\dfrac{\Delta u}{\nu_j k_0}$, ii) the turbulent intensity $\dfrac{U\sb{ns} }{ \Delta u}$, and iii) the dimensionless  cross-over scale $q_\times =\dfrac{\Omega \sb{ns} }{k_0 U\sb{ns}}=\dfrac{k_\times}{k_0}$. Here  $\Delta u$ is the root-mean-square (rms)  of the normal-fluid turbulent velocity fluctuations, $k_0=1$  is the outer scale of turbulence. The numerical values of the dimensionless counterflow velocity $V=15$ and the mutual friction frequency $\Omega=1$ and $20$ are listed in Table \ref{t:1}, columns \#8-9. In this way, for each temperature we have 4 runs, labeled below as $V_0\Omega_1$,  $V_0\Omega_{20}$, $V_{15}\Omega_1$ and  $V_{15}\Omega_{20}$.  Reference simulations of the uncoupled \Eq{NSE} with $\Omega\sb s=0, U\sb{ns}=0$, representing classical hydrodynamic turbulence for the same parameters of the flow, are labeled as "Cl".

\begin{figure*}
	\begin{tabular}{ccc}
		\includegraphics[scale=0.36]{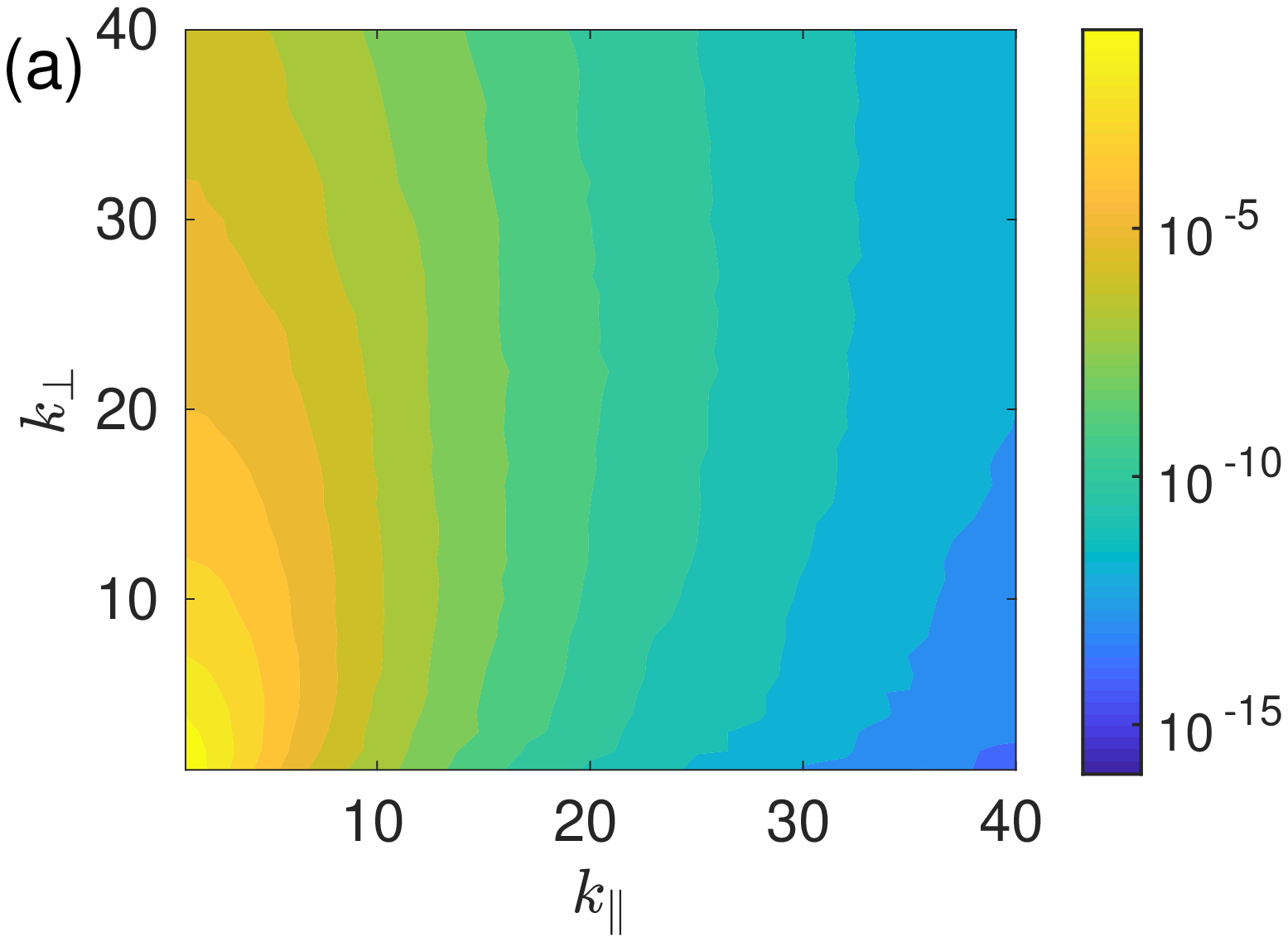}&
		\includegraphics[scale=0.35 ]{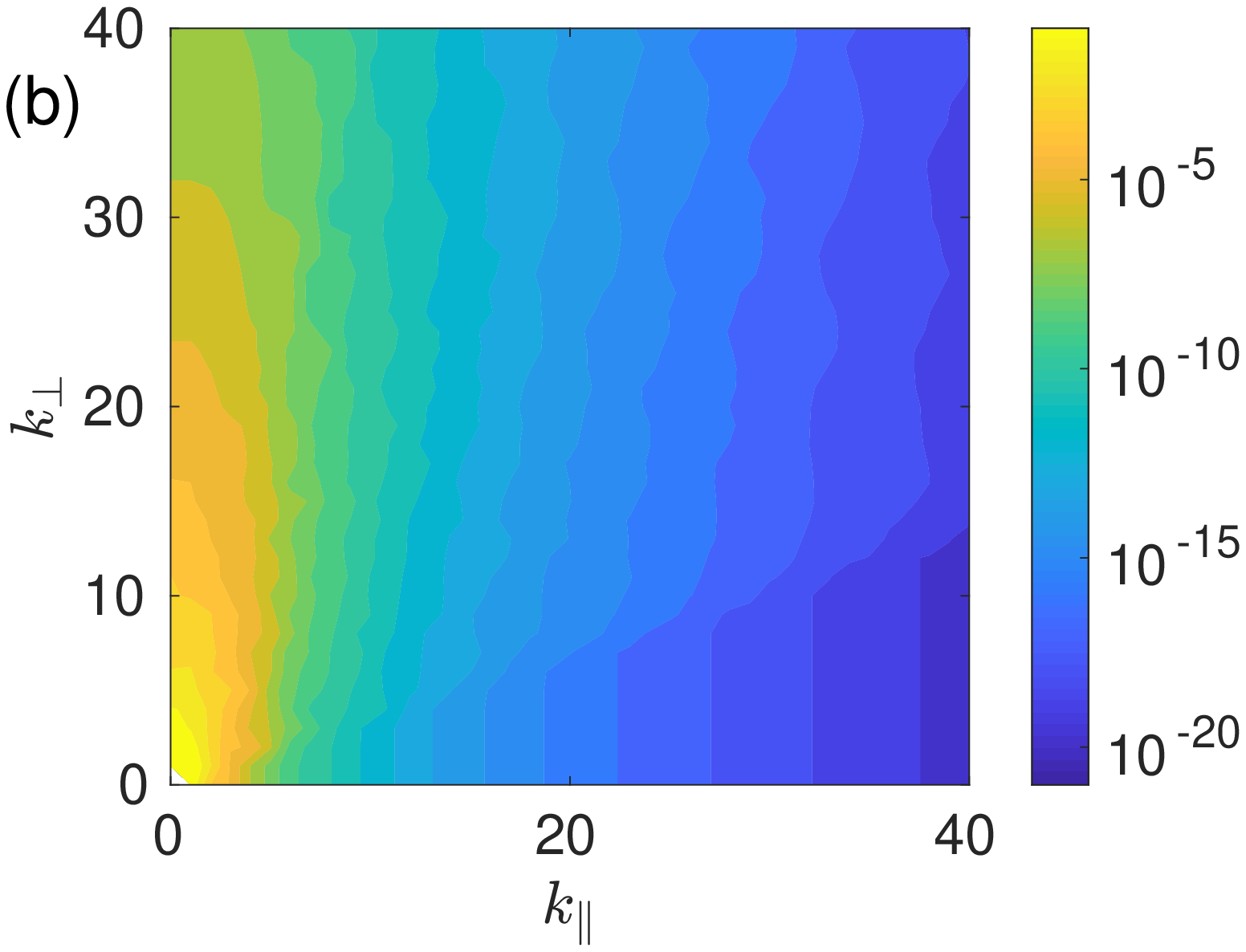}&
		\includegraphics[scale=0.36]{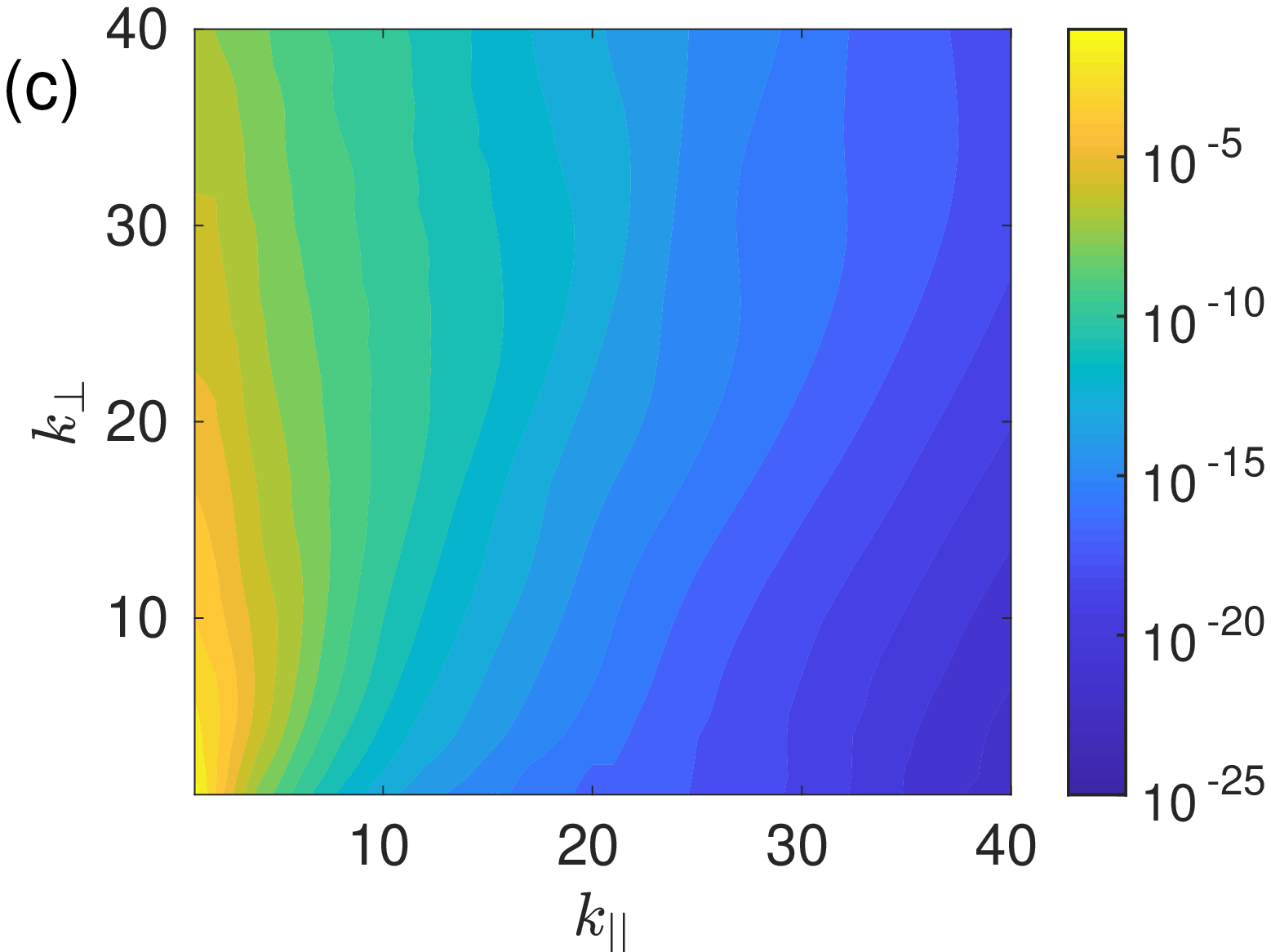}
	\end{tabular}
	\caption{ \label{f:3}The superfluid component energy spectrum
		$E\sb{s}(k_{||},k_{\bot})$  in the counterflow. (a) $T=1.65$K, (b) $T=1.85$ K, (c) $T=2.0$K. Note the difference in the magnitudes, shown by the color-bar range.}  			
\end{figure*}
The correlation between components in the counterflow  become gradually weaker
with increasing $\cos \theta$  for $T=1.65$\,K, while for $T=2.0$\,K the normal fluid and superfluid are essentially uncorrelated for  $\cos \theta\gtrsim 0.1$.

\begin{figure*}
	\begin{tabular}{ccc}
		\includegraphics[scale=0.3 ]{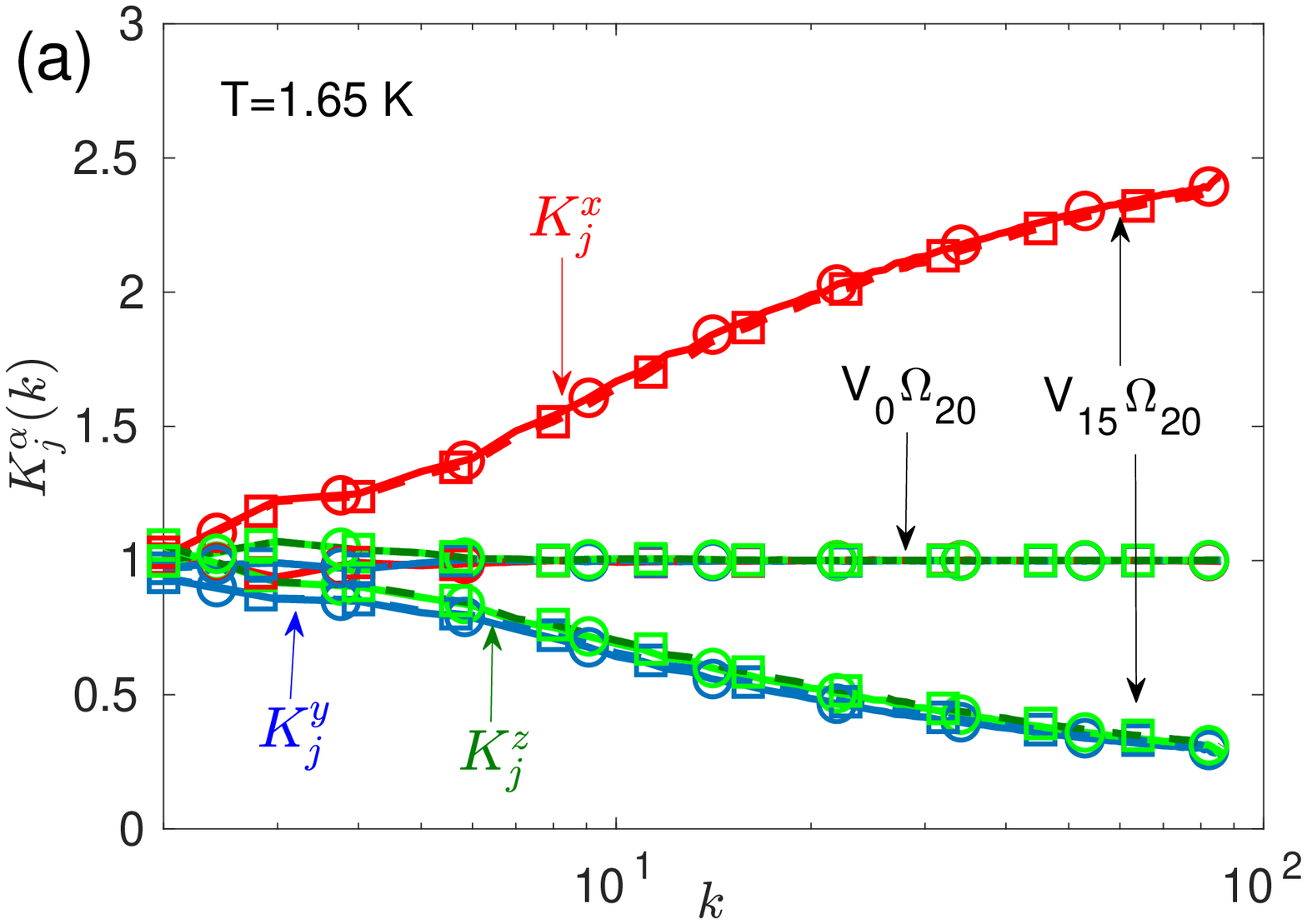}&
		\includegraphics[scale=0.3 ]{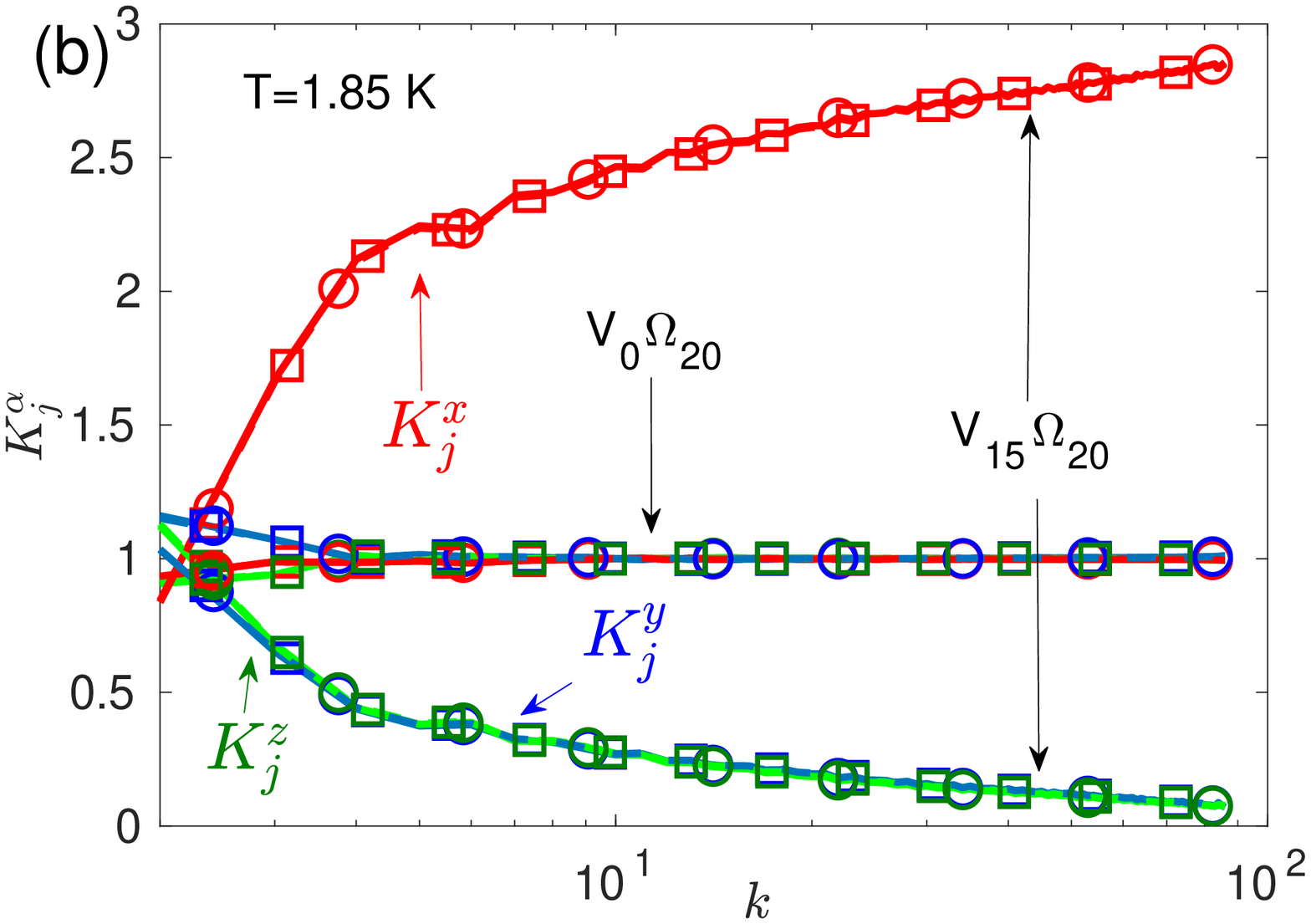}&
		\includegraphics[scale=0.3] {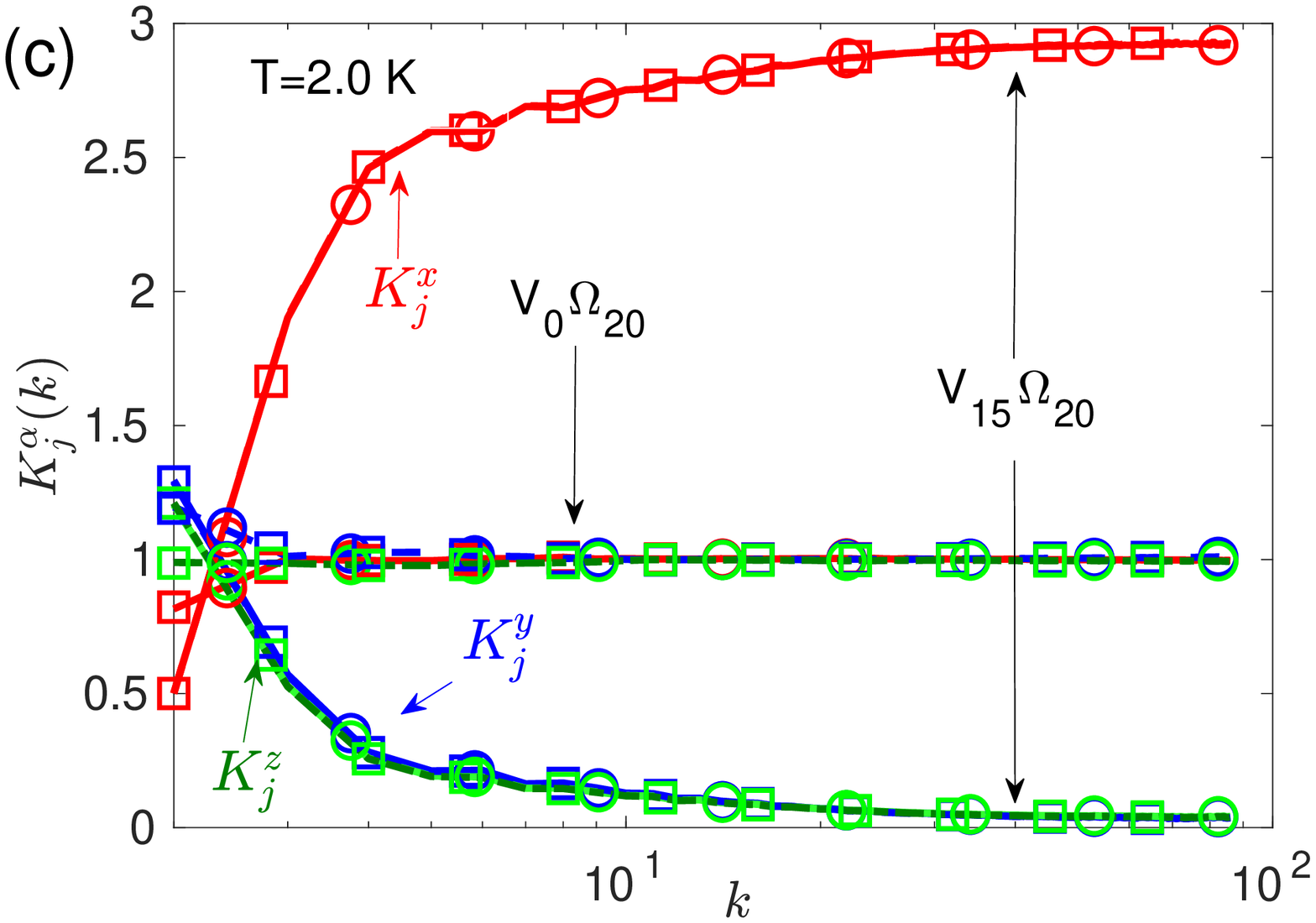}
	\end{tabular}
	\caption{ \label{f:4}
		Tensor decomposition of the 1D energy spectra $K^{\alpha}_j$ for the normal fluid (circles) and for the superfluid components (squares).
		The spectra for the coflow are labeled $V_0\Omega_{20}$, for the counterflow -- $V_{15}\Omega_{20}$. The $K^{x}_j$, $K^{y}_j$ and $K^{z}_j$ tensor components are annotated and denoted by red, blue and green lines, respectively.}  	
\end{figure*}


\subsection{\label{ss:1D-energy}  Spherical energy spectra $^{\bullet\!}E_{jj}(k)$ and cross-correlations $^{\bullet\!}E\sb{ns}(k)$}

The energy spectra are influenced by a few competing factors: the viscous dissipation, the dissipation by mutual friction and  the decoupling due to counterflow velocity.
To find their relative importance, we first ignore the present angular dependence and consider the spherically averaged spectrum $^{\bullet\!}E_{j}(k)$ and the normalized cross-correlation function:
\begin{equation}\label{R}
^{\bullet\!}R (k)  =   \frac{2\, ^{\bullet\!}E\sb{ns}(k)} { ^{\bullet\!}E\sb {nn}(k)+  \, ^{\bullet\!}E\sb {ss}(k)}\, .
\end{equation}

\begin{figure}
  	\includegraphics[scale=0.9]{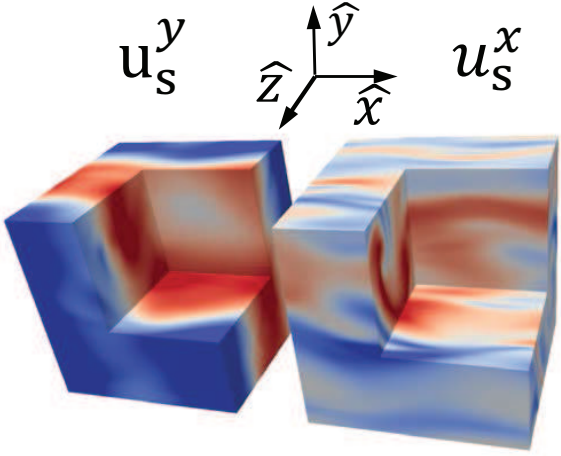}
  	\caption{\label{f:5}Superfluid velocity 	$u^x\sb s$  and 	$u^y\sb s$  components\cite{Typo}. $T=1.85\,$K,  $\Omega=20$, $V=15$. The
  		$u^z\sb s$ component (not shown) is similar to the $y$ component. The velocity
  		magnitude is color coded, with red denoting positive and blue
  		denoting negative values. }
  \end{figure}

\begin{figure*}
	\begin{tabular}{ccc}
		(a)   & (b) & (c)  \\	
		\includegraphics[scale=0.3 ]{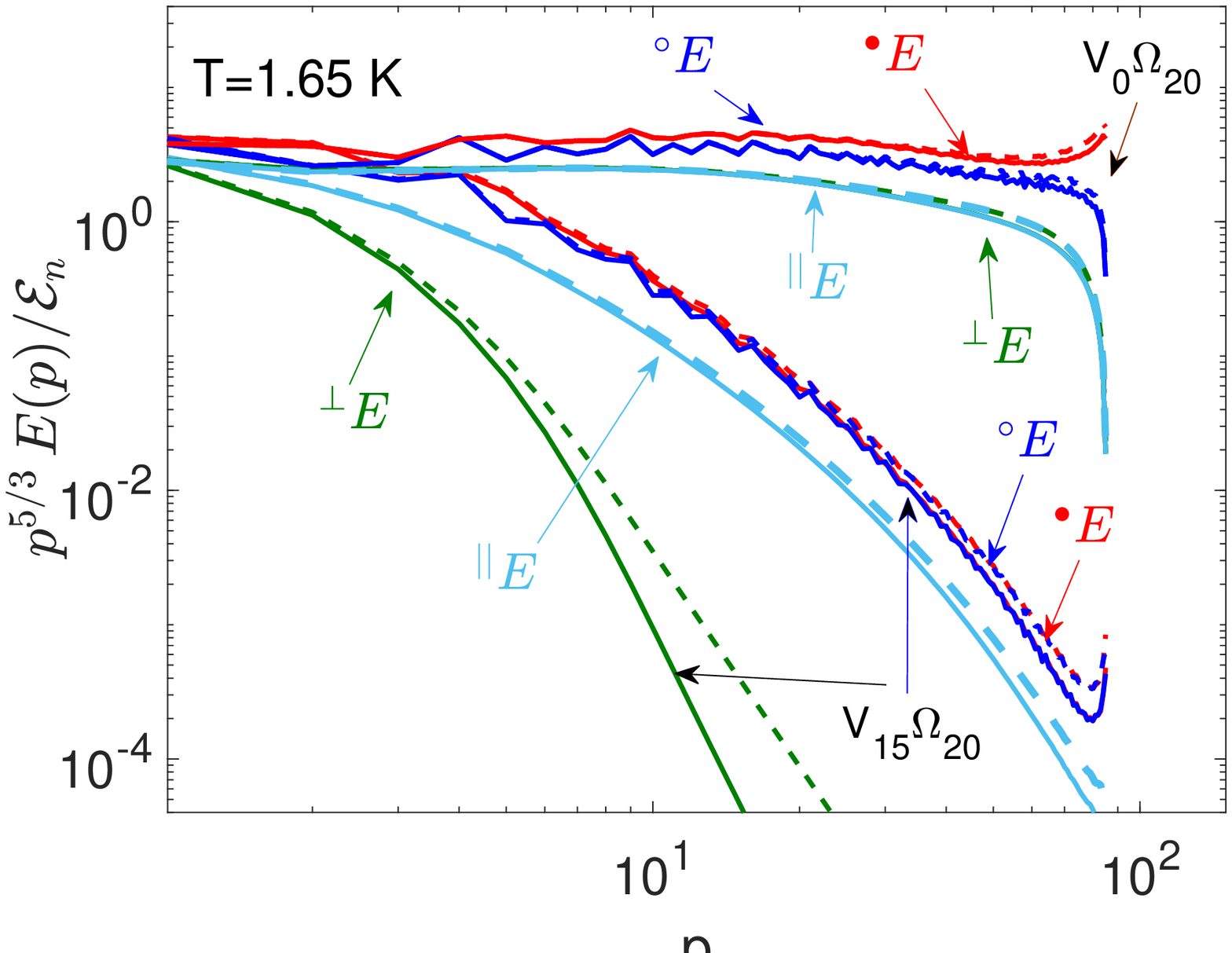}&
		\includegraphics[scale=0.3 ]{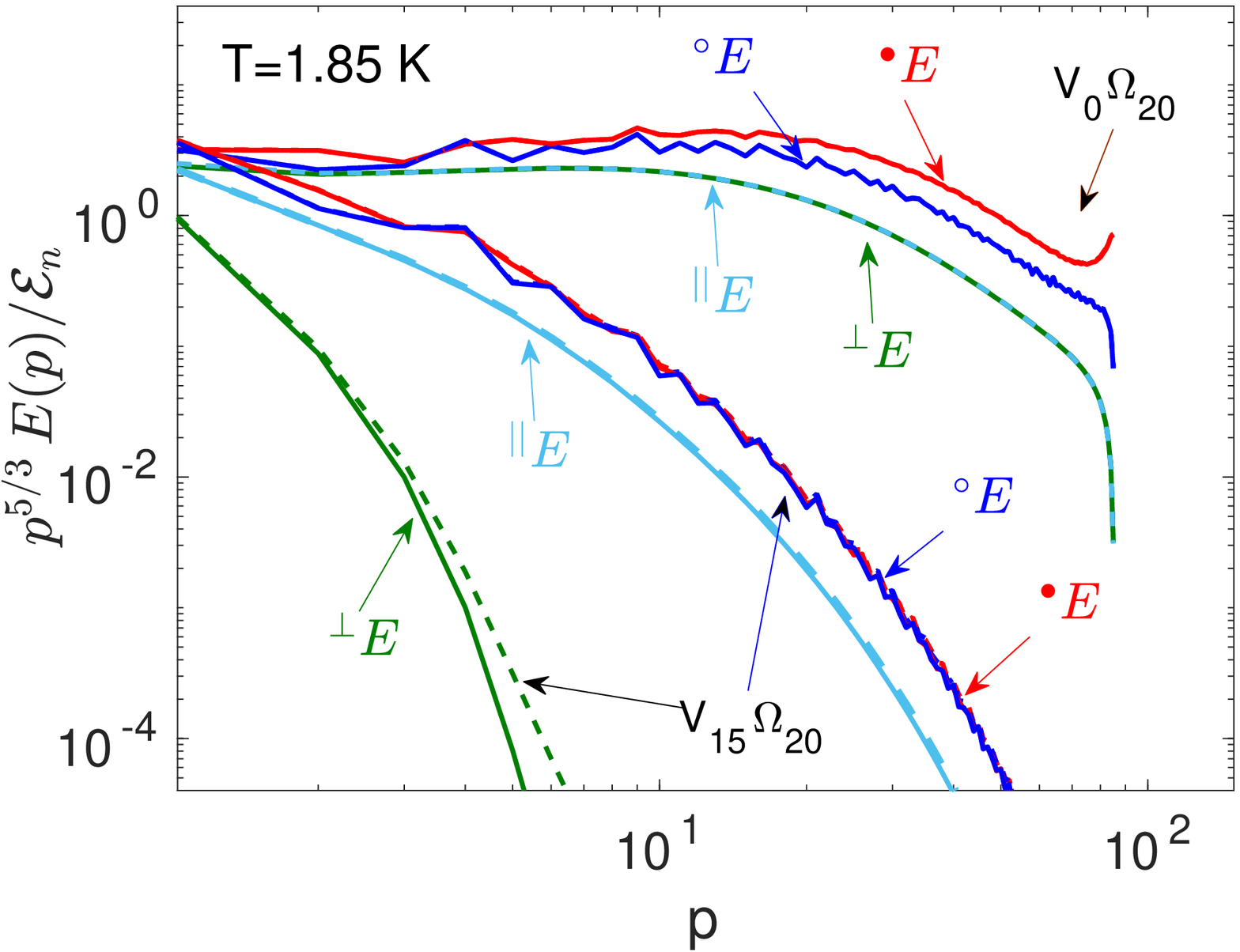}&
		\includegraphics[scale=0.3 ]{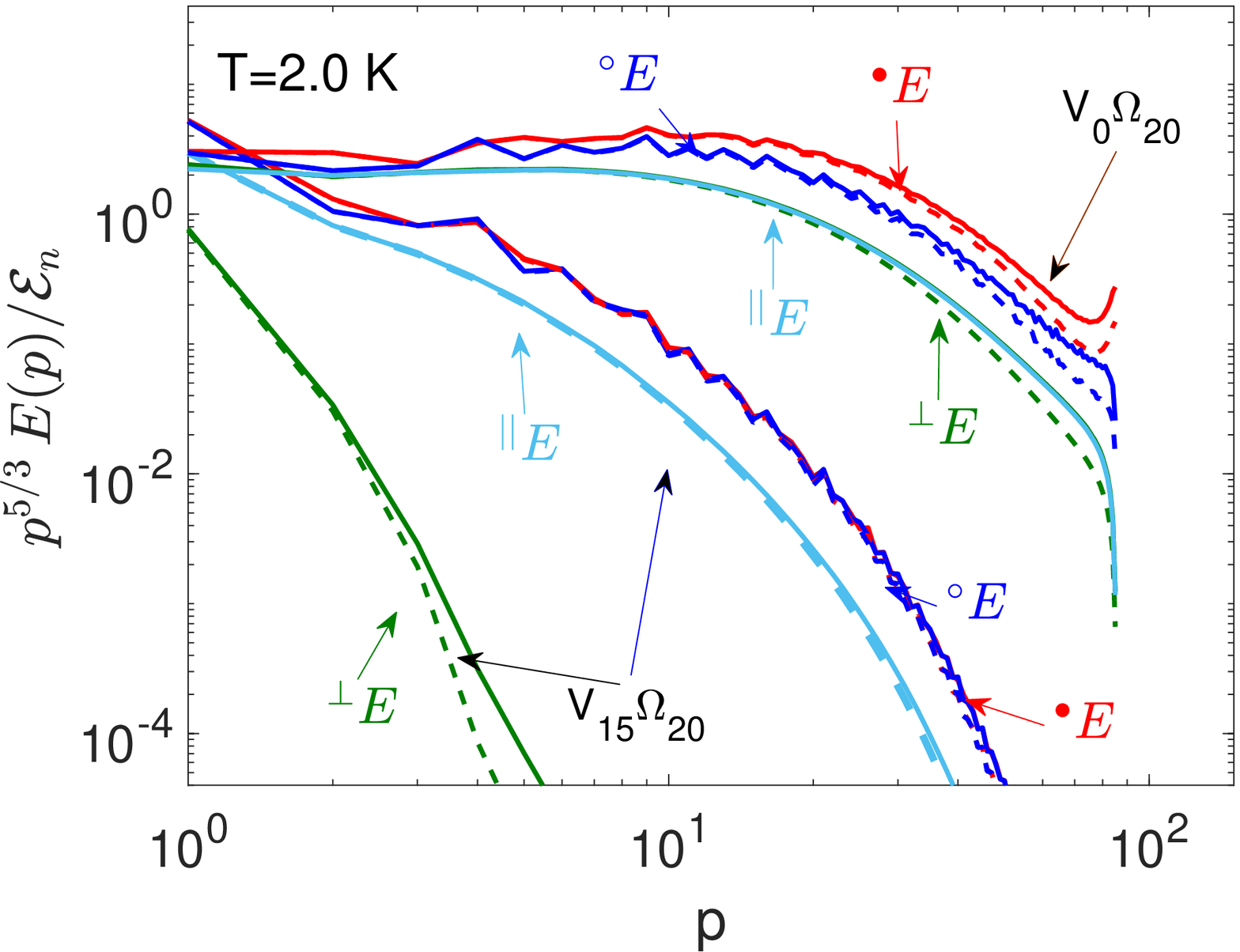}\\
		
	\end{tabular}
	\caption{\label{f:6}  Various  1D energy spectra of the normal-fluid (solid lines) and the superfluid (dashed) components for the coflow (lines labeled $V_0\Omega_{20}$)  and for the counterflow (lines labeled $V_{15}\Omega_{20}$). Different kinds of spectra are annotated in the figure.}
\end{figure*}

\begin{figure*}  	
	\includegraphics[scale=0.35  ]{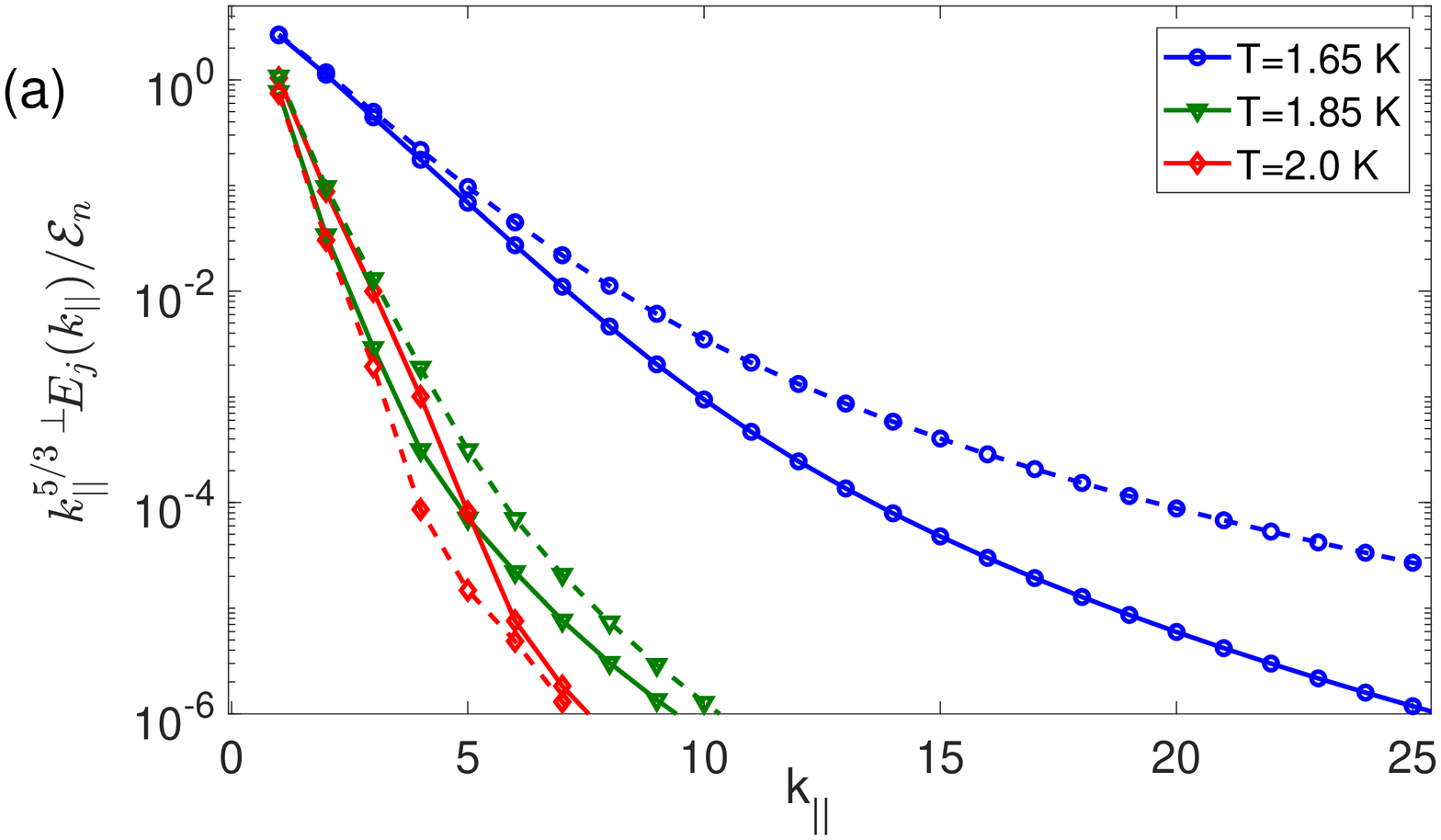}
	\hskip 1 cm 	\includegraphics[scale=0.35  ]{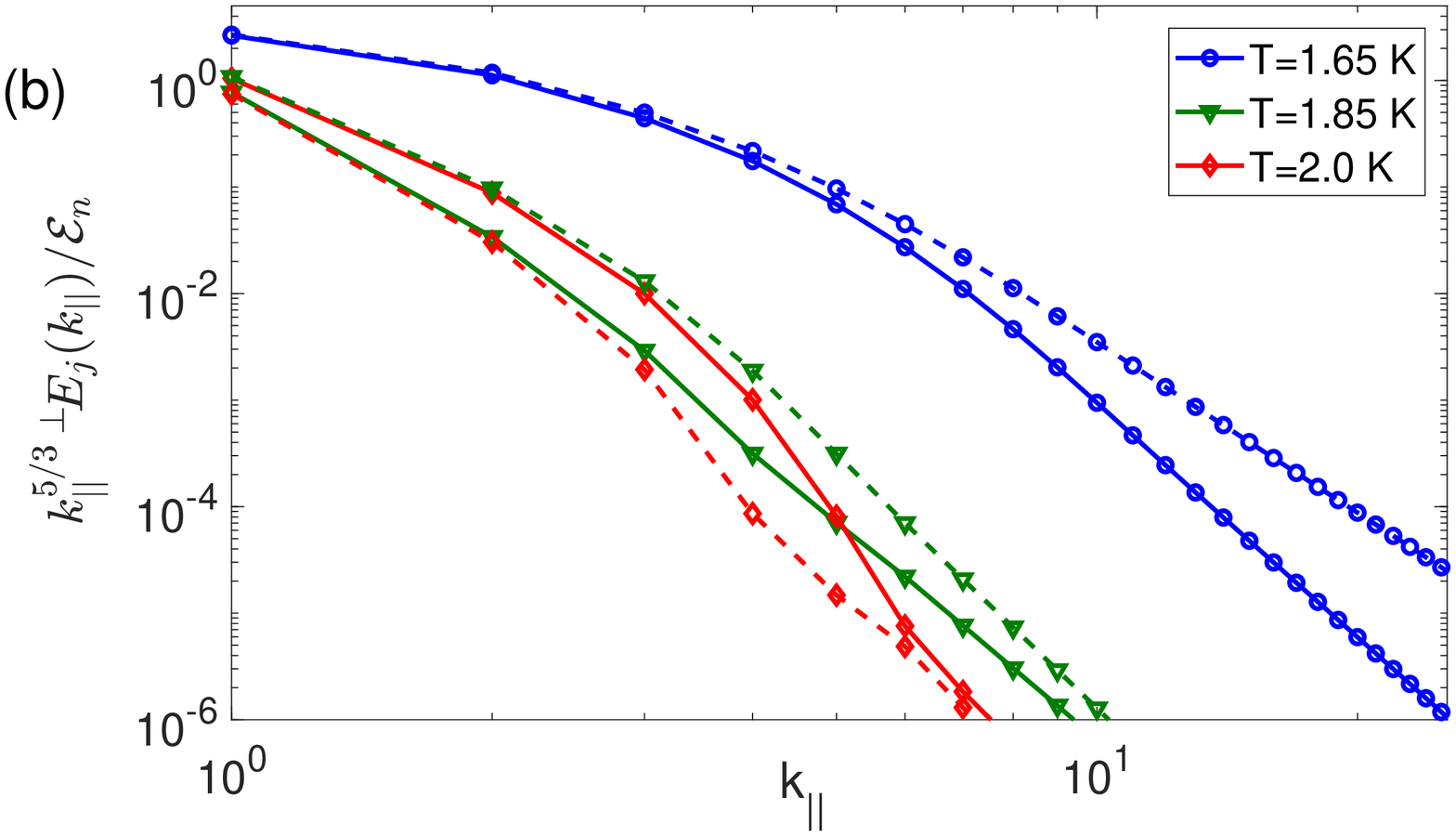}
	\caption{\label{f:7}  The  $\perp$-plane  energy   spectra $^{\perp\!}E_j(k_\|)$ of the normal-fluid (solid lines) and the superfluid (dashed) components for  counterflow $V_{15}\Omega_{20}$.  The  spectra are shown in a  Log-linear scale in panel (a)  and in Log-log scale in panel (b).  Different temperatures  are annotated in the figure. Note linear behavior at small $k_{\|}$ in panel (a) and at large $k_{\|}$ in panel (b) .}
\end{figure*}

 In the upper row of Fig.\ref{f:1} we plot the  spectrum $^{\bullet\!}E_{j}(k)$, compensated by the classical scaling $k^{5/3}$ for different flow conditions.
In the lower row we show the cross-correlations \Eq{R}. In this figure and in \Figs{f:2}-\ref{f:7} the results for $T=1.65$\,K are shown in the left column [panels (a) and (d)], for $T=1.85$\,K-- in the middle column [panels (b) and (e)] and for $T=2.0$\,K -- in the right column [panels (c) and (f)].  The effect of viscous dissipation is clearly seen in the spectra of the uncoupled components, corresponding to classical turbulence and marked "Cl", black lines. The spectra almost coincide for $T=1.85$K, for which the viscosities are almost equal. The viscosity of the normal-fluid component (solid lines) is larger than for the superfluid (dashed lines) for $T=1.65$\,K and smaller for $T=2.0$\,K.

 Next, we add the coupling by the mutual friction force, creating a coflow (green and brown lines). The strongly coupled components ($V_0\Omega_{20}$, brown lines) are well correlated at all scales and move almost as one fluid. The corresponding spectra slightly differ only at the viscous scales. Note the additional dissipation due to mutual friction, leading to further suppression of the spectra compared to the uncoupled case, for $T=1.85$ and $2.0$\,K. At the lower temperature $T=1.65$K the energy exchange between components leads to stronger  dissipation in the superfluid component and weaker dissipation in the normal-fluid component. For weaker coupling ($V_0\Omega_{1}$, green lines), the situation is completely different. The components are almost uncorrelated, especially at large $k$. The coupling between them is translated into very efficient dissipation by mutual friction, leading to spectra that are suppressed almost at all scales, especially at high temperature (see \Ref{DNS-He4} for details).

In the presence of the counterflow velocity\cite{decoupling},  the two components are swept in opposite directions by the corresponding mean velocities. This leads to further decorrelation of the component's turbulent velocities, especially at small scales, for which the overlapping time is very short [cf. lines for $V_{15}\Omega_{1}$ and $V_{15}\Omega_{20}$ in \Fig{f:1}(d-f)]. Even for the strong coupling, $V_{15}\Omega_{20}$, blue lines, the velocities become progressively less correlated for all temperatures. The dissipation by mutual friction is very strong in this case, with both $\Omega$ and the velocity difference being large, leading to very strongly suppressed spectra, with $ \, ^{\bullet\!}E\sb {nn} (k)\approx  \, ^{\bullet\!}E\sb {ss}(k)$. At $T=1.65$K there is still some interval of scales with $k\gtrsim k_0$, for which the spectra are close to K41 scaling. The crossover scale agrees well with $k_{\times}\approx 7$ for this case (see Table \ref{t:1}, column \#14). For higher temperatures this crossover scale become smaller and the classical-like behavior is not resolved.  At weak coupling ($V_{15}\Omega_{1}$, red lines), the velocities are essentially uncorrelated and the spectra of the two components differ and are very strongly suppressed, especially at high $T$.

\subsection{\label{ss:2D} Angular dependence of 2D-energy spectra}
The behavior of the spherically averaged energy spectra agrees well with the predictions of the theory\cite{LP-2018}, based on the assumption of spectral isotropy.  To explore the angular dependence of the energy spectra and the correlations $\tilde E_{ij}(k,\theta)$ we plot in \Fig{f:2}(a-c) the spectra  $\tilde E_{j}(k,\theta)$, normalized by the corresponding $^{\bullet\!}E_{j}(k)$  and in \Fig{f:2}(d-f) the corresponding normalized cross-correlations $\tilde R(k,\theta)$.
Given the discrete nature of the $\B k$-space in DNS, we further average them over 3 bands of wavenumbers. We do not account for the largest scales $k\approx k_0$ which are influenced by the forcing and average the spectra and the cross-correlations over the $k$-ranges $10\leq k< 20$,  $20\leq k< 60$  and   $60\le k \le 80$. The corresponding lines are labeled  as $k_{10}$, $k_{20}$ and $k_{60}$, respectively.
Here we consider only strong coupling regime and plot the spectra  and the cross-correlations for the coflow ($V_0\Omega_{20}$) and for the counterflow  ($V_{15}\Omega_{20}$) .

The first observation is that the spectra and the cross-correlation for the coflow are isotropic for all the conditions. The angular dependencies of $\tilde E_j(k, \theta)$ and $\tilde R(k, \theta)$ for the counterflow, on  the other hand, have a complicated form. Both the spectra and cross-correlation are largest for $\cos\theta\approx 0$ and fall off very quickly with decreasing angle. The spectra decrease exponentially with $\cos\theta$, slower for small  $k$ (red lines for $k_{10}$) and faster for larger $k$ (green and blue lines for $k_{20}$ and $k_{60}$, respectively). This effect is stronger for the normal-fluid (super fluid) component at low temperatures (high temperature). Most of the energy is contained in the narrow range $\cos \theta <0.1$,  near the $\perp$-plane in the $\B k$-space,  orthogonal to $\B U\sb{ns}$.

To better quantify the angular energy distribution, we use the fact that the spectra  $\tilde E_{j}(k, \theta)$  have piecewise exponential dependence of $\cos\theta$ , as is evident from \Fig{f:2}(a)-(c). We then estimate the $\cos\theta$-range, in which half of the total energy is contained, for different wavenumber bands. At $T=1.65$\,K, for the small wavenumbers $k_{10}$ band, this range is indeed $\cos(\theta)<0.1$  for both the normal an superfluid components. With increasing temperature, this range decreases to  $\cos \theta <0.05$ for  $T=1.85$\,K and to $\cos \theta<0.03$ for the normal-fluid  and $\cos \theta<0.025$ for the superfluid at  $T=2.0$\,K. For the larger $k_{20}$ band these values are $0.04$ and $0.06$ for the normal fluid and the superfluid, respectively, at $T=1.65$\,K. For higher temperatures, as well as for the high wavenumber $k_{60}$ band, about a half of the total energy is contained in a narrow $\cos \theta<0.02-0.025$ range for both components.

Indeed, the superfluid energy spectrum $E\sb{ss}(k_{||},k_{\bot})$, shown in \Fig{f:3}, is strongly suppressed in $k_{||}$ direction, while decreasing slowly  in the orthogonal direction, especially for $T=2.0$ K.


  \subsection{\label{ss:tens} Tensor structure of 1D energy spectra}
  Given such a strong anisotropy of the spectra in the counterflow, it is natural to expect that different components of the turbulent velocity fluctuations are excited to a different extent. In this section we consider the  tensor  structure of 1D-energy spectra $^{\bullet\!}E^{\alpha\alpha}_{jj}(k)$ for $\alpha=x,\, y,\, z$ and clarify which components ($v_j^x$ along $\B U\sb{ns}$ or $v_j^y,v_j^z$ , both orthogonal to $\B U\sb{ns}$) are most excited.

  In  \Figs{f:4} we plot the components of the spherical spectra for three temperatures as the ratios
  \begin{equation}\label{tensor_ratio}
  ^{\bullet\!}K_j^\alpha(k) \= 3\, ^{\bullet\!}E^{\alpha\alpha}_{jj}(k)/\, ^{\bullet\!}E_{jj}(k) \ .
  \end{equation}
  The factor 3 was introduced to ensure that for the isotropic turbulence  $\displaystyle \sum_{\alpha=x,y,z}\,   ^{\bullet\!} K _j^\alpha(k)=1$.

  Indeed, for the coflow  (the almost horizontal lines, labeled $V_0\Omega_{20}$) all the velocity components are excited equally, except for the smallest wavenumbers.
On the other hand, for the counterflow turbulence (lines labeled $V_{15}\Omega_{20}$) the contribution of the  $ ^{\bullet\!}K_j^x(k)$ component (shown by red lines) is dominant and monotonically increasing with $k$ from the isotropic level $ ^{\bullet\!}K_j^x(k_0)\approx 1$ to the maximal possible level $ ^{\bullet\!}K_j^x(k )\approx 3$. This means that the small-scale counterflow turbulence mainly consists of $v_j^x(\B k)$ velocity fluctuations. The contribution of $v_j^y$ and $v_j^z$ fluctuations for $k\gtrsim 10$ is negligible, especially at $T=2.0$\,K.

 Therefore, the counterflow turbulence represent a special kind of a quasi-2D turbulence, consisting mostly of the turbulent  velocity fluctuations with only one stream-wise projection $\B u_\|$, which depends on the cross-stream coordinate $\B r_\perp$: $\B u_\|(\B r_\perp, t)$.
	This behavior is essentially different from other known types of quasi-2D turbulence, such as stably-stratified flow in the atmosphere\cite{atmosTurb-Kumar,2018-AB,A5} or rotational turbulence
 \cite{rot1,rot2,rot3},  in which  the leading contribution to the turbulent velocity field comes   from the 2D velocity field $\B u_\perp$ that depends   on  $\B r_\perp$:   $\B u_\perp(\B r_\perp, t)$.
Such a $\B u_\|(\B r_\perp, t)$-turbulence can be visually presented as narrow jets or thin sheets
as illustrated in \Fig{f:5}.
	
Note the difference with the strong acoustic turbulence. There the velocity field has tangential velocity breaks at the jets boundaries and the 1D energy spectrum $E(k)\propto k^{-2}$. The energy spectra in the counterflow decay much faster. It means that the velocity fields  at the jets boundaries are continuous together with some finite number of their derivatives. This is a consequence of the mutual friction that tends to smooth the velocity field.

\subsection{\label{ss:comp}   Comparison of 1D energy    spectra and reconstruction of 3D spectra}

The best way to study the anisotropy of hydrodynamic turbulence is to expand the statistical objects in the irreducible representations of the	SO(3) symmetry group, see, e.g. \Refs{A1,A2,A3,A4,A5}. In  counterflow turbulence, an attempt to expand   $\tilde E_j(k,\theta)$  into a series with respect to Legendre polynomials,  \begin{equation}\label{Pol}
	\tilde E_j(k,\theta)=\sum _\ell ~ \tilde E_j(k,\ell)\, P_\ell (\cos\theta)\,,
	\end{equation}
	 and to study $k$-behavior of $ \tilde E_j(k,\theta)$, turned out to be ineffective. The very strong anisotropy  of $\tilde E_j(k, \theta)$ spectra required too many terms in the expansion\,\eqref{Pol} for an adequate reproduction of its angular dependence. Therefore, we choose another way to characterize the spectral anisotropy, which is more suitable in our case. We compare the  normal-fluid  and superfluid  spherical $^{\bullet\!}E_j(k)$, cylinder $^{\circ\!}E_j(k_{\perp})$, and  $\|$-, $\perp$-plane-averaged energy spectra $^{\|\!}E_j(k_{\perp})$, $^{\perp\!}E_j(k_{\|})$.
	
	 In the case of isotropy, all the four  1D energy spectra are proportional to each other	
	 \begin{equation}\label{Esp2a}
	 		^{\bullet\!}E_{ij }^{\alpha\beta} (p )\propto\,  ^{\circ\!}E_{ij }^{\alpha\beta} (p)\propto\, ^{\perp\!}E_{ij }^{\alpha\beta} (p)\,\propto\,  ^{\|\!}E_{ij }^{\alpha\beta} (p)\, ,
	 		\end{equation}
	differing only in numerical prefactors. Here $p$ is the corresponding (dimensional, $[p]=$1/cm) wavenumber: $p=k,\ k_\perp,\ k_\|$ or $k_y$. By estimating contributions to the integrals in  \Eqs{Esp}, \eqref{Esp1E} and \eqref{Esp1D} in the case of strong anisotropy (i.e coming from a narrow range with $ k_\| \ll k_\perp $), one may show that the spectra
	   are related as $^{\bullet\!}E_j(p) \approx \, ^{\circ\!}E_j(p) \approx C \,  ^{\|\!}E_j(p)$, where $C$ is a numerical prefactor.
	  This fact may explain the good agreement between the experimental spectra $^{\|\!}E\sb n(k_{\bot})$, obtained in \Ref{WG-2018} and the prediction of the theory\cite{LP-2018} for $^{\bullet\!}E_j(k)$.
	    The integral in \Eq{Esp1C} is different and the spectrum $^{\perp\!}E_j(k_{\|})$ is expected to be confined to small $k_{\|}$  range.

 These spectra, normalized by the energy density $\C E\sb n$ and  compensated by the K41 factor  $p^{5/3}$,  are shown in \Fig{f:6}.
 The  coflow spectra, appearing as almost horizontal lines, labeled $V_0\Omega_{20}$, indeed differ by less than an order of magnitude for all $T$.  The relation between various spectra for the counterlow is consistent with the above estimate, further confirming the strong spectral anisotropy. The degree to which the $^{\perp\!}E_j(p)$  spectra, shown by green lines, are suppressed at different temperatures,  agrees with the angular dependence of $\tilde E_j(k,\theta)$, \Fig{f:2}(a)-(c). While at $T=1.65$\,K the spectrum for the $k_{10}$-range at  $\cos \theta \approx 1$  is smaller by three orders of magnitude  in the direction of the counterflow compared to the orthogonal plane,  at $T=2.0$\, K this difference is almost ten orders of magnitude. Accordingly, the $^{\perp\!}E_j(p)$  spectrum at $T=2.0$\, K is confined to less than a decade in $p$.
To better quantify the steepness of the spectra we list in Table \ref{t:2} the values of the ratios $E_j(10)/E_j(1)$ for   $^{\perp\!}E_j(p)$  and  $^{||}E_j(p)$.

 \begin{table}
 \begin{tabular}{c|c|c|c}
 	\hline \hline
   &$T=1.65$K &  $T=1.85$K&  $T=2.0$K\\
 	\hline
 $^{\|} E\sb {n,s}(10)/\,  ^{\|} E\sb {n,s}(1)$ & $10^{-3}$& $2.0\times 10^{-4}$ & $2.5\times 10^{-4}$  \\
 	\hline
   $^{\perp\!} E\sb n(10)/\,  ^{\perp\!} E\sb n(1)$& $6.3\times 10^{-6}$  & $2.0 \times 10^{-8}$ &  $1.6\times 10^{-8}$\\
 	\hline
 	 $^{\perp\!} E\sb s(10)/\,  ^{\perp\!} E\sb s(1)$& $2.5\times 10^{-5}$  & $3.2 \times 10^{-8}$ &  $1.6\times 10^{-9}$\\
 	\hline \hline
 \end{tabular}
 	\caption{\label{t:2} The steepness of the energy spectra  $^\perp E_j(k)$ and  $^\| E_j(k)$ in the counterflow, characterized by the ratios $^{\perp\!} E_j(10)/\,  ^{\perp\!} E_j(1)$ and $^{\|\!} E_j(10)/\,  ^{\|\!} E_j(1)$.}
 \end{table}
\begin{figure*}
	\begin{tabular}{ccc}
		(a)   & (b) & (c)  \\	
		\includegraphics[scale=0.3 ]{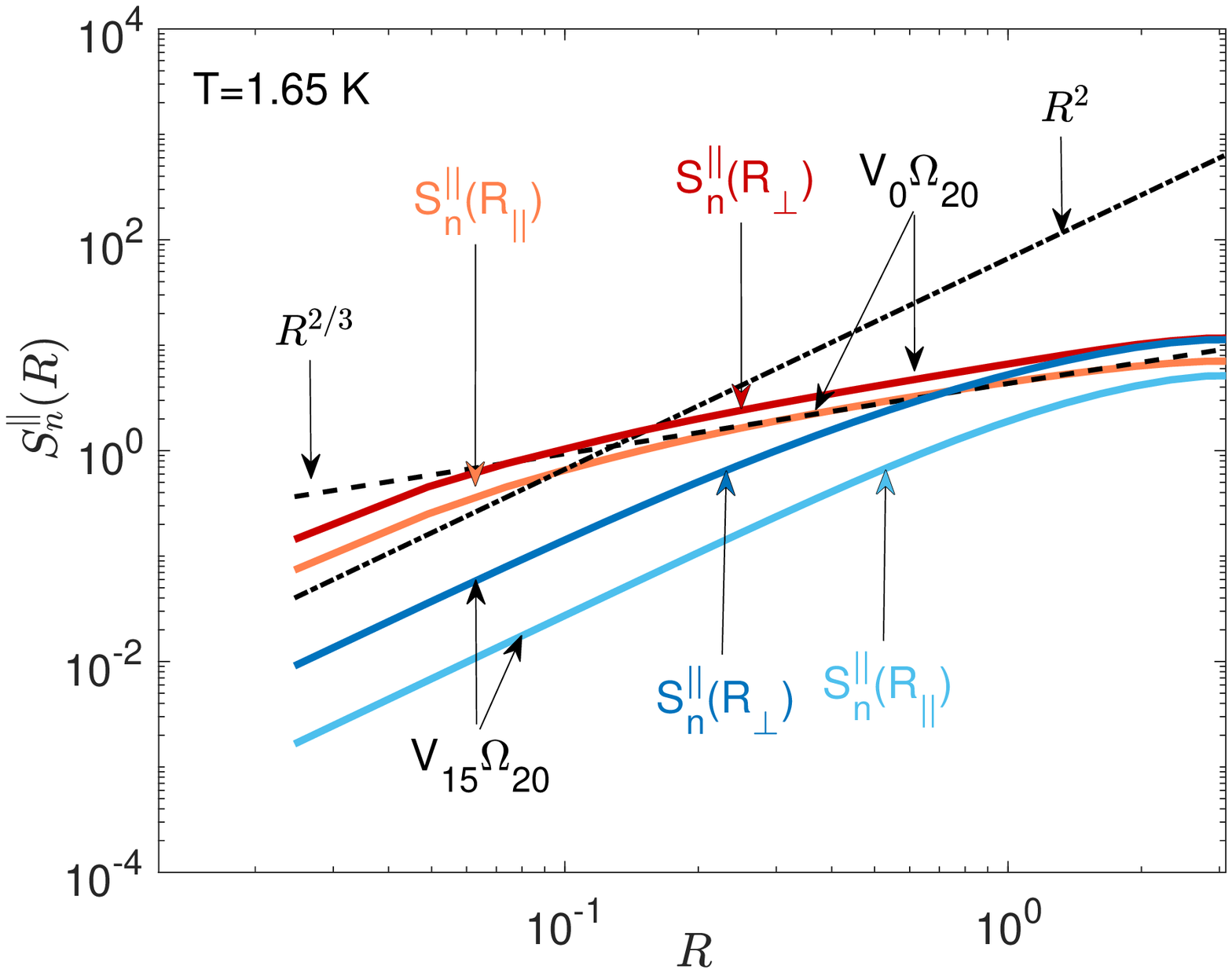}&
		\includegraphics[scale=0.3 ]{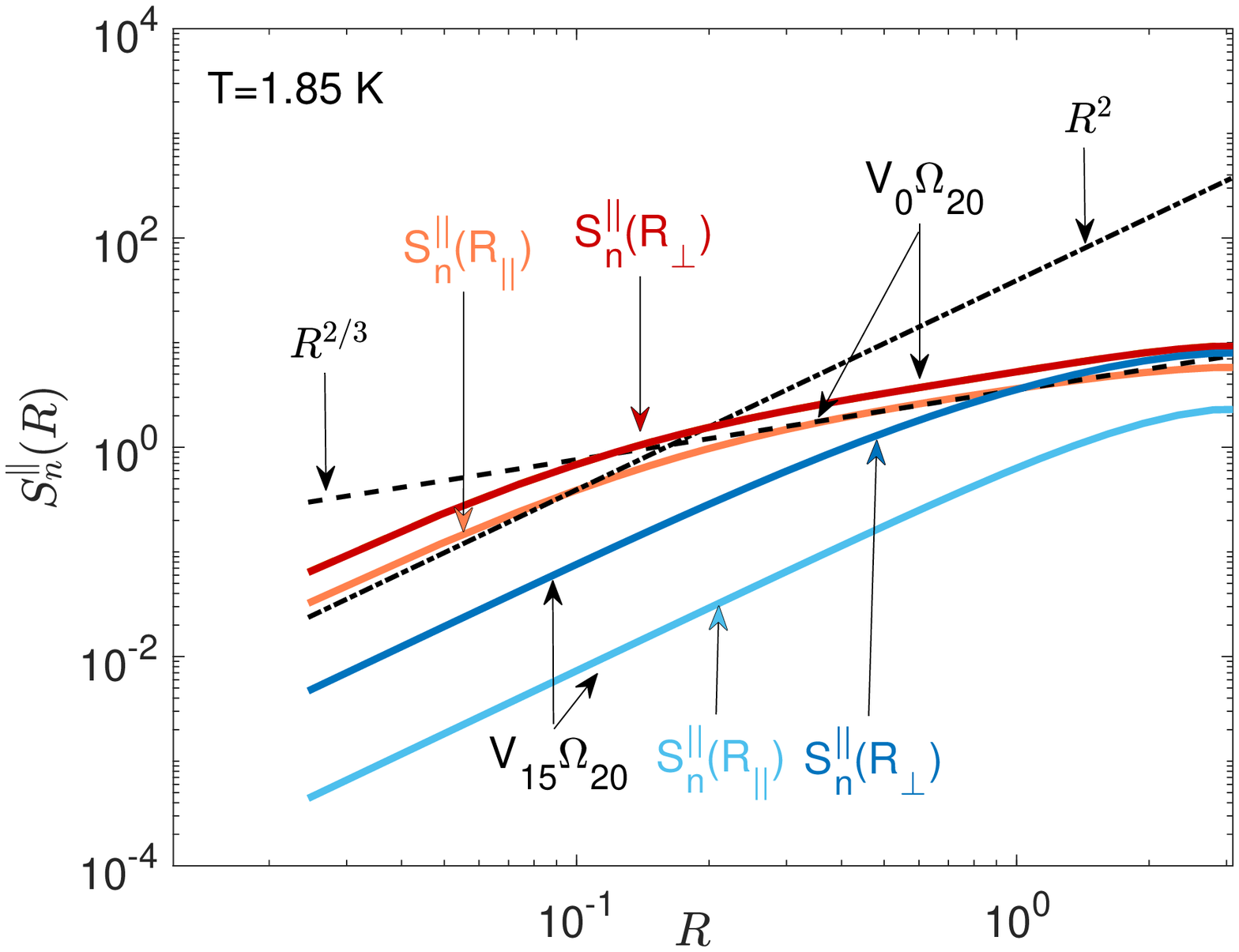}&
		\includegraphics[scale=0.3 ]{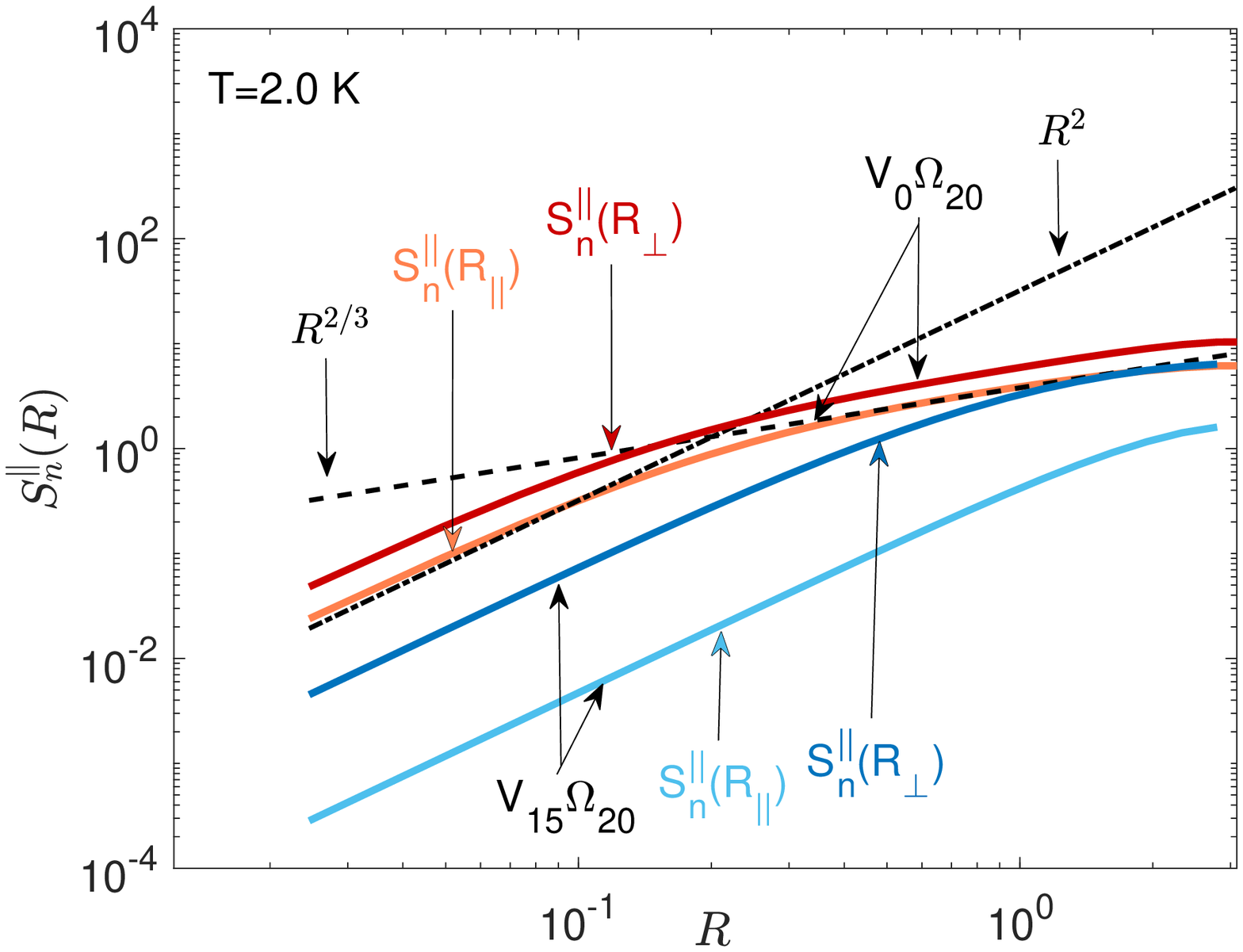}\\		
	\end{tabular}
	\caption{\label{f:8} The velocity structure functions of the normal-fluid component $S\sb n^{||}(R)$ \eqref{S2b}.  The lines for the coflow are marked $V_{0}\Omega_{20}$, for the counterflow $V_{15}\Omega_{20}$. Various structure functions are marked in the figures. The dashed lines, labeled  $R^{2/3}$ and the dot-dashed lines, marked $R^2$, serve to guide the eye only.}
\end{figure*}
\begin{figure*}  	
	\includegraphics[scale=0.35  ]{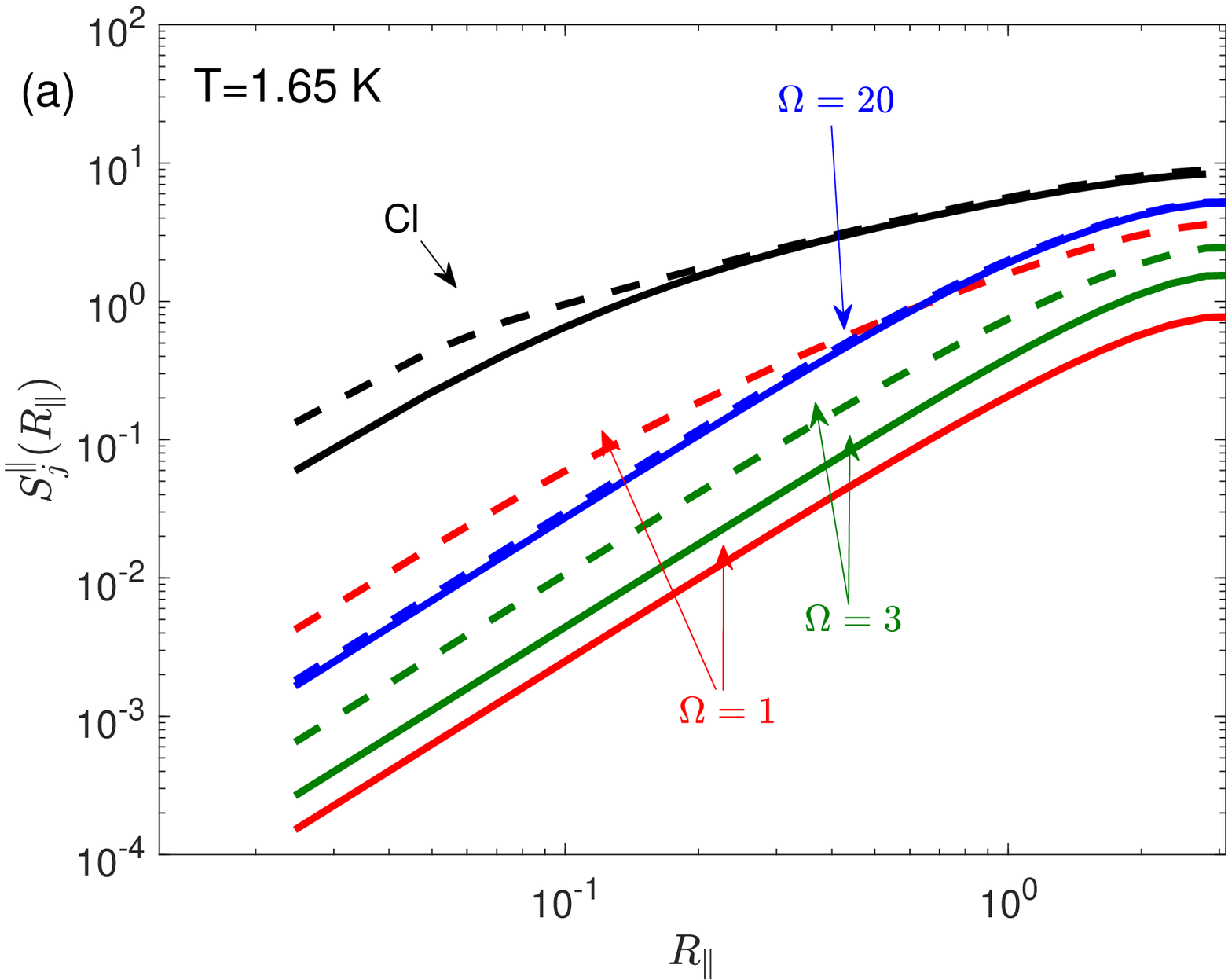}
	\hskip 1 cm 	\includegraphics[scale=0.35  ]{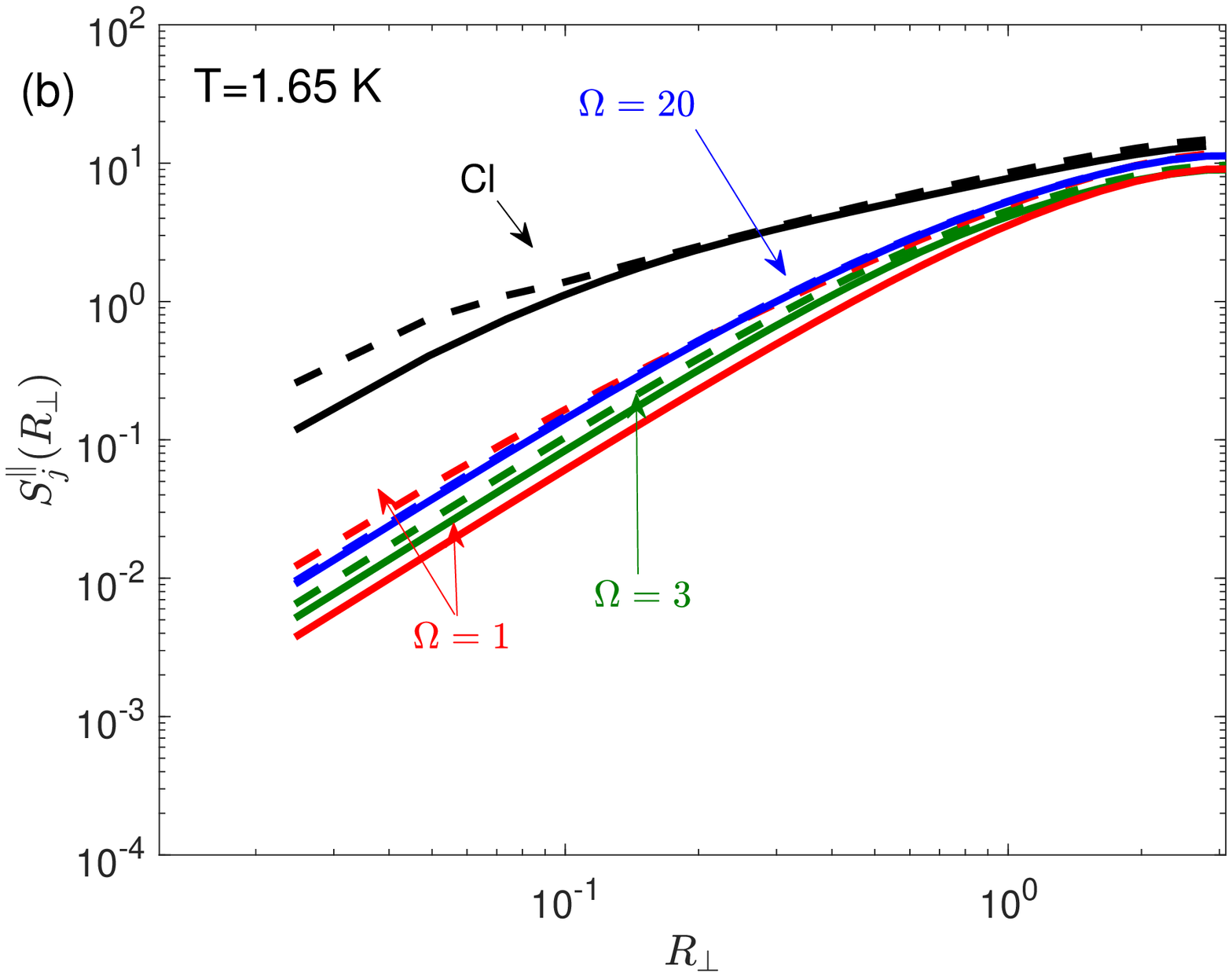}
	\caption{\label{f:9}  The structure functions $S_j^{||}(R_{||})$ (a) and $S_j^{||}(R_{\bot})$ (b) at $T=1.65$ K. The lines for the counterflow with $V=15$ and different values of $\Omega$ are color-coded: $\Omega=1$ (red lines), $\Omega=3$ ( green lines), $\Omega=20$ ( blue lines). The structure functions for the normal-fluid component are shown by solid lines and for the superfluid component--by dashed lines. Structure functions for the classical turbulence ($V=0, \Omega=0$) are shown by black lines and labeled "Cl". }
\end{figure*}
\begin{figure}  	
	\includegraphics[scale=0.4 ]{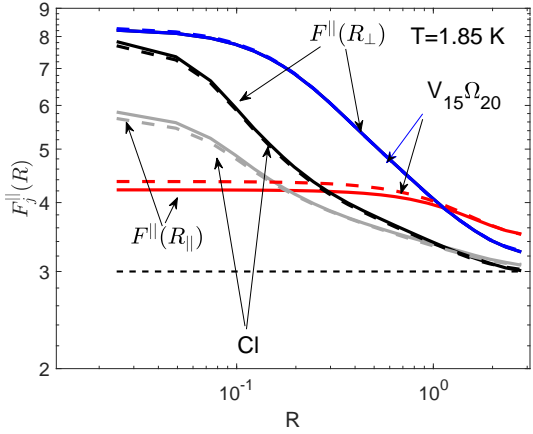}
	\caption{\label{f:10}  The flatness $F_j^{||}(R_{||})$ (red lines)  and $F_j^{||}(R_{\bot})$ (blue lines) $T=1.85$ K in the counterflow $\Omega\sb s=20, V=15$.   The flatness for the normal-fluid component are shown by solid lines and for the superfluid component--by dashed lines. The corresponding flatnesses for the classical turbulence ($V=0, \Omega=0$) are shown by black and gray lines and labeled "Cl".}
\end{figure}
\begin{figure*}
	\begin{tabular}{ccc}
		(a)   & (b) & (c)  \\	
		\includegraphics[scale=0.3 ]{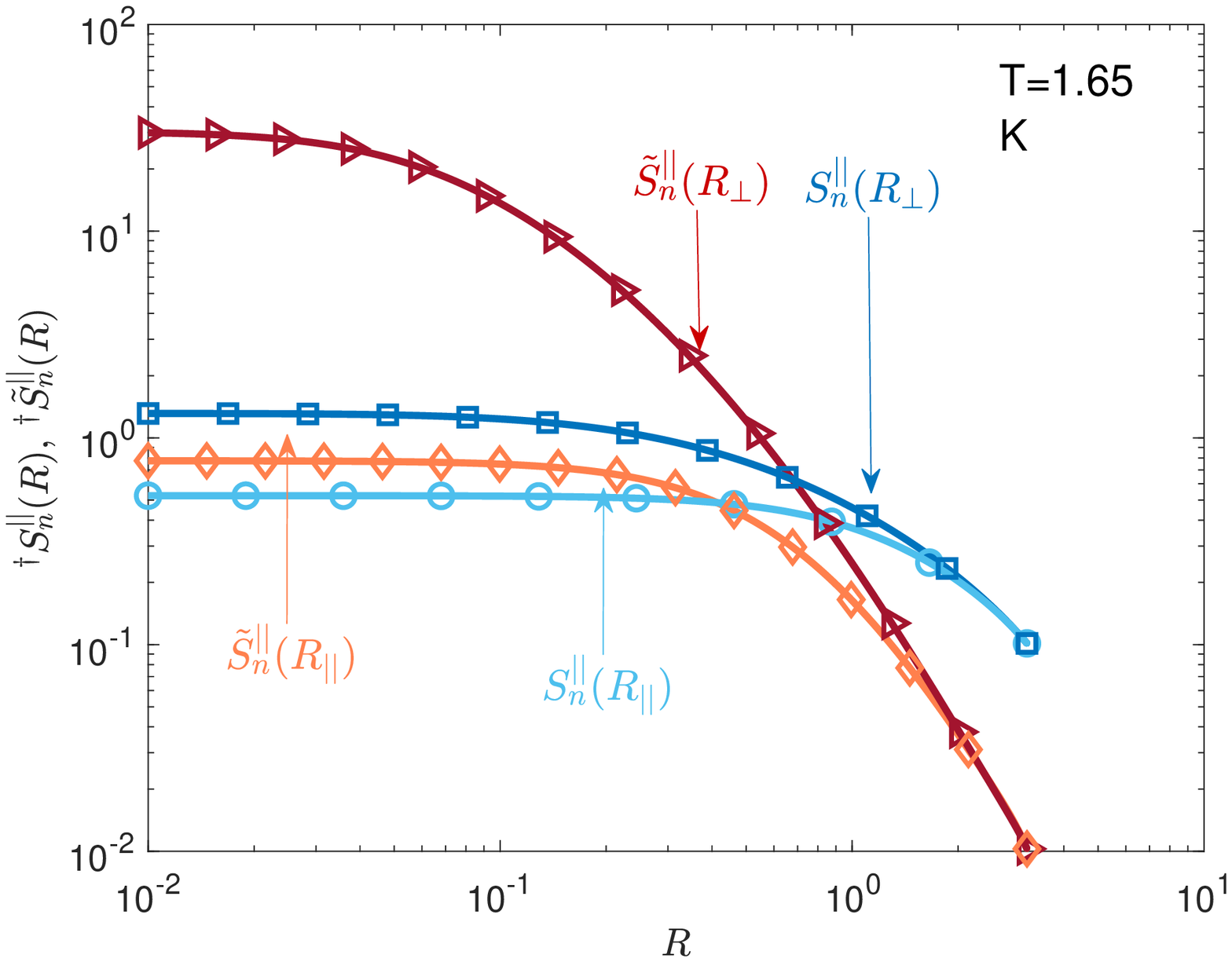}&
		\includegraphics[scale=0.3 ]{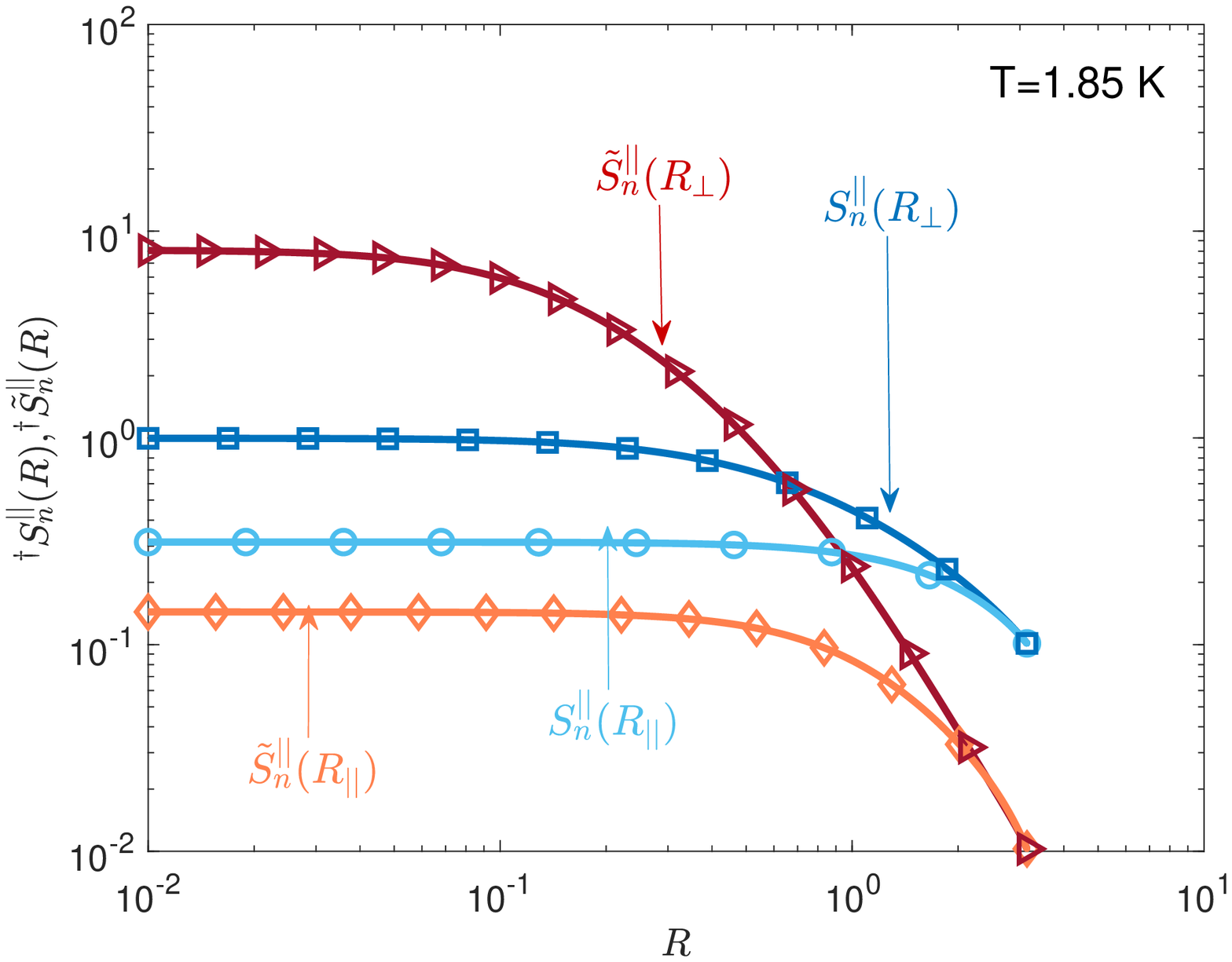}&
		\includegraphics[scale=0.3 ]{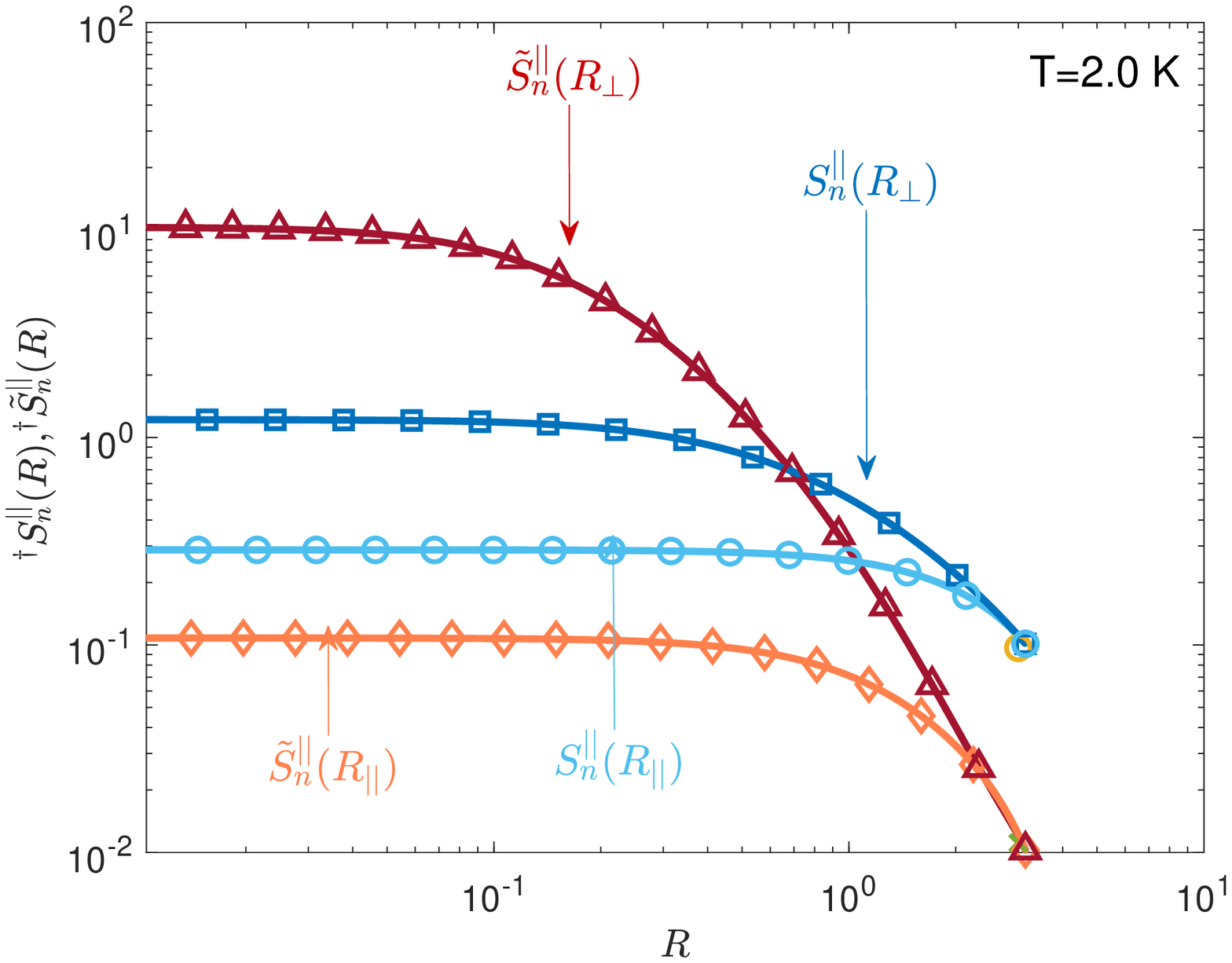}\\		
	\end{tabular}
	\caption{\label{f:11} Comparison of the  normalized compensated velocity structure functions $^{\dagger}S\sb n^{||}(R_{\perp})$ ($\Box$) and $^{\dagger}S\sb n^{||}(R_{\|})$ ($\bigcirc$)  with $^{\dagger}  \tilde S\sb n^{||}(R_{\perp})$($\rhd$) and  $^{\dagger}  \tilde S\sb n^{||}(R_{\|})$ ($\Diamond$) for the normal-fluid component in the counterflow.}
\end{figure*}

 The analysis of the $(k,\theta)$-dependence of the  $\tilde E_j(k,\theta)$  energy spectra in \Sec{ss:2D} showed that the  overwhelming part of the total turbulent energy is concentrated in the range of small $\cos\theta$, and consequently small $k_\|$, say for $k_\|\lesssim 10$. 	For a semi-qualitative analysis of the 2D-spectra $E_j(k_\|,k_\perp)$ and $\tilde E_j(k,\theta)$   in this range of  $k_{||}$, we assume a factorization
			\begin{subequations}\label{fact}
				\begin{equation}\label{factA}
			E(k_\|,k_\perp)\simeq \, f_1 (k_\|)\,  f_2  (k_\perp)\ .
				\end{equation}
			If so, using \Eqs{Esp1E} and \eqref{Esp1C}, we can reconstruct the 2D energy spectra as follows
				\begin{equation}\label{factB}
			E_j(k_\|,k_\perp) \simeq  \frac{\,   ^{\circ \!}E_j(k_\perp)  \, ^{\perp \!} E_j(k_\|)}{    E_j}\,,
				\end{equation}
			where $  E_j$ is the   energy density in the system given by \Eq{Esp2B}.
			
			Furthermore, using \Eqs{def-F}, we can also reconstruct the 2D spectra $\tilde E(k ,\theta)$ from $E(k_\|,k_\perp)$:
		\begin{equation}\label{factC}
		\tilde E_j(k,\theta) \simeq  \frac{\,   ^{\circ \!}E_j(k\sin\theta)  \, ^{\perp \!} E_j(k\cos\theta)}{  E_j \sin\theta  }\ .
		\end{equation}
	\end{subequations}

 In the range  of small $\cos \theta$, where most of the turbulent energy is concentrated, \Eq{factC} can be simplified as follows: $\tilde E_j(k,\theta) \simeq   \,   ^{\circ \!}E_j(k)  \, ^{\perp \!} E_j(k\cos\theta)/ E_j$. The  $\theta$-dependence of  $\tilde E_j(k,\theta)$ is therefore determined by   $^{\perp \!} E_j(k\cos\theta)$, i.e. $\cos \theta$ appears in the combination $k\cos\theta=k_\|$. This observation fully  agrees with our theoretical prediction, that $\cos\theta$ appears in the theory only via \Eq{LP-20C} in the dimensionless factor $k U\sb{ns}\cos\theta/\Omega\sb{ns}$. We consider  this agreement as an argument in a favor of the factorization assumption\,\eqref{factA} for small $\cos\theta$.
	
 To take a closer look at $^{\perp \!} E_j(k_\|)$, we plot  in \Fig{f:7} these spectra for different temperatures.  To expose the functional dependence of the spectra, we  use different scales in two panels: in the panel (a) the scale is Log-Linear, while in panel (b) the spectra are plotted in the Log-Log scale. At all temperatures the small-$k_{\|}$ behavior is exponential, while at larger $k_{\|}$ the spectra are consistent with the power-law behavior. 	
	Using this information, we propose the following form for the small-$k_{\|}$-spectra:
\begin{equation}\label{estE}
 ^{\perp\!}E_j(k_\|)\simeq \frac{k_\|\,\C E_j}{k_*}\, \exp \Big[-\frac{k_\|}{k_*}\Big ]  \ .
\end{equation}
It is tempting to relate the characteristic $k_*$ to the crossover scale $k_\times$ : $k_*\propto \Omega\sb{ns}/U\sb{ns}=k_\times$. Indeed,  $k_*$ estimated from  \Fig{f:7}(a)  and $k_\times$ (Table\,\ref{t:1}, column \#\,14) have similar temperature trends. This gives additional support for factorization\,\eqref{factA} and for qualitative theoretical discussion of the problem in \Sec{ss:theory}.

 The  observed steep power-law behavior of $^{\perp\!}E_j(k_\|)$ for larger   $k_{\|}\gtrsim 10$,  \Fig{f:7}(b) with an apparent exponents $m\approx -7$ may  indicate  a nonlocal energy transfer between largest  and smallest scales, similar to the super-critical spectra in the superfluid $^3$He.

 \subsection{\label{ss:SF}  The structure functions}
The energy spectra $^{\|\!}E_j(k_{\perp})$, $^{\perp\!}E_j(k_{\|})$ may be  translated into the corresponding structure functions, according to \Eqs{S2-1} and \eqref{TS3-1}. In \Fig{f:8} we show the structure functions \eqref{S2b}  $S_n^{||}(R)\equiv S\sb n^{xx}(R)$ with the velocity differences taken in the direction of the counterflow $R_{||}\equiv R_x$ and in the plane $R_{\bot}$ orthogonal to it.
The structure functions \eqref{S2b} for the coflow, shown in \Fig{f:8}(a-c) as  red and orange lines are similar  to classical turbulence; at large scales they follow approximately $R^{2/3}$ scaling, gradually crossing over towards  viscous $R^2$ behavior. The transition is very broad  here, but the two apparent scaling ranges are evident.  The cross-over scale increases with temperature. The structure functions, calculated along and across the counterflow direction, are similar, slightly differing mostly in the magnitude at all scales. The main difference from the uncoupled case (not shown) is the lower magnitude at all scales, reflecting the presence of addition energy dissipation by mutual friction.
In the counterflow, the situation is different. Over  most of the available range of  scales, the structure functions, shown as dark and light blue lines in \Fig{f:8}, appear to have an apparent scaling behavior close to $R^2$, especially $S\sb n^{||}(R_{||})$. The actual behavior depends on the flow conditions, in agreement with the results of \Ref{WG-2017}. The magnitudes of the structure functions are much lower than for the coflow. At the lower temperature $T=1.65$K, $S\sb n^{||}(R{\bot})$ has an overlap with the corresponding structure function in the coflow a large scales, which disappears with increasing temperature. The two types of the structure functions in the counterflow have significant difference in magnitude, with  $S\sb n^{||}(R_{||})$ being strongly suppressed. As it was suggested in Sec.\ref{ss:VSF}, these structure functions do not quantitatively reflect the corresponding energy spectra, however the qualitative difference should be observable experimentally.

The influence of the coupling strength on the behavior of the structure functions  is illustrated in \Fig{f:9} for $T=1.65$K. Here, in addition to the weak coupling $\Omega=1$ and the strong coupling $\Omega=20$ we consider also an intermediate coupling strength $\Omega=3.4$. The structure functions for the classical turbulence are included for comparison.  The general form is similar for all values of $\Omega$, with all the structure functions in the counterflow being strongly suppressed compared to classical turbulence, especially $S_j^{||}(R_{||})$. Note that at this temperature the structure functions of the normal fluid are more suppressed for {\it weaker} coupling, in accordance with the energy spectra in \Fig{f:1}(a). For the transverse $S_j^{||}(R_{\bot})$ the difference between the two fluid components is relatively small and the influence of the coupling strength is weak. This is consistent with the 2D energy spectra,  shown in \Fig{f:3}: the energy spectra in the transverse direction $k_{\bot}$  are weakly influenced by the mutual friction.

Additional information may be obtained from analysis of the flatness $F_j^{||}(R)=P_j^{||}(R)/[S_j^{||}(R)]^2$, where $P_j^{||}(R)=\langle (\delta_{\B R} u^{x}_j)^4) \rangle$ 	is the forth-order structure function.  In \Fig{f:10} we compare $F\sb n^{||}(R_{\bot})$ and $F\sb n^{||}(R_{||})$ with the flatness in the classical turbulence. In the transverse direction, $F\sb n^{||}(R_{\bot})$ in the counterflow is growing towards small scales faster than in the classical turbulence at large scales. This indicate a moderate enhancement of intermittency at intermediate scales in this direction, in agreement with experimental results of \Ref{WG-2018}.  In the longitudinal direction, similar to the structure functions, the flatness $F\sb n^{||}(R_{||})$ is almost constant in the counterflow, leading to a much stronger discrepancy between the transverse and the longitudinal components that in the classical turbulence. This constant value reflects the behavior of the  structure functions $S\sb n^{||}(R_{||})\propto R_{||}^2$ and  $P\sb n^{||}(R_{||})\propto R_{||}^4$ over a wide range of scales that is a consequence of 
 the energy spectra $^{\perp\!}E_j(k_{\|})$ that fall off faster than $k^{-3}$. 

The second difference structure functions
 $\tilde S^{||}_j(R)$ \eqref{S3b} are expected to better reflect the underlying spectra, at least for $^{\|}E(k_\perp)$, since they have wider windows of locality up to $k^{-5}$.   The steeper  $^{\perp \!} E(k_\|)$ result in $S^{||}(R_{\|})$ behaving as $R^4$ in most of the available range of scales.
In principle,
 these exponents can fall within the windows of locality
 for the structure functions of the third difference (up
 to $x \le 7$) and of the fourth difference (up to $x \le 9.0$).
 We do not discuss these objects due to the increasing
 difficulties in their measurements.

Having in mind possible experiments, we compare in \Fig{f:11}  two types of the structure functions for the normal fluid in the counterflow. To allow a meaningful comparison we plot them normalized by the values at the largest $R=R\sb{max}$ and compensated by the corresponding viscous scaling
 \begin{eqnarray}
 ^{\dagger}S^{||}_j(R)&=&S^{||}_j(R)/S^{||}_j(R\sb{max}) R^{-2}\, ,\\\nonumber
  ^{\dagger}\tilde S^{||}_j(R)&=&\tilde S^{||}_j(R)/\tilde S^{||}_j(R\sb{max}) R^{-4}\ .
 \end{eqnarray}
 	Indeed, the transition to the viscous behavior  (the horizontal lines at small scales) occurs at smaller $R$ for  $\tilde S^{||}_j(R)$ (marked by triangles and diamonds) than for $S^{||}_j(R)$ (marked by squares and circles). As expected, the range of the condition-dependent apparent  scaling at large scales also increases. In addition, the difference in the amplitudes of the structure functions  in the  longitudinal and transverse directions is much larger  for $\tilde S^{||}(R)$, hopefully allowing more accurate detection of the anisotropy.

\section{\label{s:con}Conclusions}

Both the theoretical considerations and the results of the numerical simulations presented indicate strong
 anisotropy in the energy distribution in  counterflow turbulence.
This is basically due to an angular dependence of the energy dissipation caused by the mutual friction force. It  tends to suppress the velocity fluctuations elongated across the direction of the counterflow velocity. At the same time,  most of the flow energy is confined to a narrow wavenumber plane, orthogonal to this direction, leading to a flow which is smooth along the counterflow direction and turbulent across it. Unlike rotational and atmospheric turbulence with stable stratification, in counterflow turbulence  the streamwise velocity component  plays the dominant role. This effect is progressively stronger at smaller scales and at higher temperatures. At low temperatures, the milder gradual increase of the small scale anisotropy  is due to the smaller fraction of normal fluid and consequently weaker decorrelation.
The structure functions of this anisotropic, non-scale invariant  turbulent flow, do not allow to extract the quantitative information
about the energy distribution over scales, but are
expected to  reveal strong differences between the directions along and orthogonal to the counterflow velocity.
\acknowledgments
L.B. Thanks Michele Buzzicotti for the data analysis and visualization. GS thanks AtMath collaboration at University of Helsinki.


\begin{thebibliography}{99}
\bibitem{Typo} Please notice that in the \Ref{AnisoLetter} there is a typo referring to these data as $u_{\rm s,x}^2$ and $u_{\rm s, y}^2$.
\end{thebibliography}
\end{document}